\documentclass{elsart}
\usepackage[square]{natbib}

\usepackage{amssymb}

\usepackage[dvips]{graphicx}
\usepackage{psfrag}
\journal{Astroparticle Physics}

\begin{document}

\begin{frontmatter}



\title{Monte Carlo simulations of geosynchrotron radio emission from CORSIKA-simulated air showers}


\author[FZKIK]{T. Huege\corauthref{cor}},
\corauth[cor]{Corresponding author.}
\ead{tim.huege@ik.fzk.de}
\author[FZKIK]{R. Ulrich},
\author[FZKIK]{R. Engel}

\address[FZKIK]{Institut f\"ur Kernphysik, Forschungszentrum Karlsruhe, Postfach 3640, 76021 Karlsruhe, Germany}

\begin{abstract}
We present simulations performed with REAS2, a new Monte Carlo code for the calculation of geosynchrotron radio emission from extensive air showers. The code uses thoroughly tested time-domain radio emission routines in conjunction with a realistic air shower model based on per-shower multi-dimensional CORSIKA-generated histograms. We assess in detail how the transition from simpler, parametrised, to realistic, CORSIKA-based particle distributions affects the predicted radio emission from a typical $10^{17}$~eV air shower. The effects of eliminating a previously needed free parameter and adopting realistic electron to positron ratios are also discussed. Compared with earlier calculations based on parametrised showers, REAS2 simulations predict slightly weaker and in some cases narrower pulses. In addition, a pronounced east-west versus north-south asymmetry arises in the emission pattern, and the radio pulses become generally unipolar. Finally, we demonstrate how REAS2 can be used to study radio pulse shapes and their relation to shower characteristics such as the longitudinal air shower development.
\end{abstract}


\end{frontmatter}

% main text
%________________________________________________________________

\section{Introduction}

In the last few years, radio detection of cosmic ray air showers has once again become a very active field of research. A number of new experiments such as LOPES \citep{FalckeNature2005,HornefferArena2005,ApelAschBadea2006,HuegeCris2006} and CODALEMA \citep{ArdouinBelletoileCharrier2005,ArdouinBelletoileCharrier2006} have been established. These experimental activities are accompanied by theoretical studies of the emission processes to build the necessary basis for the interpretation of experimental data.

There have already been a number of pioneering studies of radio emission from extensive air showers (EAS) in the context of geomagnetic and \v{C}erenkov-like mechanisms (see \citep{Allan1971} for a review). A new perspective on radio emission from EAS was introduced with the proposal to interpret the dominant geomagnetic contribution as synchrotron radiation from secondary shower electrons and positrons gyrating in the earth's magnetic field \citep{FalckeGorham2003}, hereafter called ``geosynchrotron radiation''.

Analytical calculations along that line have been presented in \citep{SuprunGorhamRosner2003} and, in association with the LOPES project, in \citep{HuegeFalcke2003a}. In the model of Huege \& Falcke \citep{HuegeFalcke2003a}, the longitudinal evolution of the air shower as well as the lateral, arrival time and energy distributions of the particles are described with well-known analytical parametrisations. The radio emission is calculated in the frequency-domain. Analytical calculations like these help in understanding complex coherence effects arising from air shower structures on length scales comparable to the observing wavelength. On the other hand, analytical calculation in the frequency-domain necessitate certain approximations such as adopting infinite particle tracks and the adoption of the Fraunhofer limit. A recent study \citep{Luo2006} takes into account the effect of semi-infinite particle tracks. A general drawback of analytical calculations, however, remains: they are very limited in the description of the extremely complex air shower characteristics.

A natural choice to tackle this complexity is to revert to Monte Carlo simulations. In \citep{HuegeFalcke2005a} we have described a Monte Carlo code for the calculation of geosynchrotron {\bf r}adio {\bf e}mission from {\bf a}ir {\bf s}howers (REAS1). The code calculates the emission in the time-domain. This approach has several advantages: no far-field approximations have to be made, the finiteness of the particle trajectories is taken into account in a natural way, and polarisation characteristics of the emission can be easily calculated. In order to allow a good comparison with the results of \citep{HuegeFalcke2003a}, the air shower model was kept identical, i.e., the air shower evolution and particle distributions were described with the same analytical parametrisations as in \citep{HuegeFalcke2003a}. This code allowed extensive analyses of the relationship between air shower parameters and the associated radio emission, the results of which have been published in \citep{HuegeFalcke2005b}.

Other efforts to simulate radio emission from cosmic ray air showers with Monte Carlo techniques have recently been undertaken. Essentially the same formalism that was used in \citep{HuegeFalcke2005a} has been incorporated into the AIRES \citep{Sciutto1999} code for extensive air shower simulations. Some first results presented in \citep{DuVernoisIcrc2005} showed qualitatively similar results to those published in \citep{HuegeFalcke2005b}. In a somewhat different approach, the EGSnrc \citep{EGSnrc} code has been adapted to make frequency-domain radio calculations taking into account the refractive index profile of the atmosphere, i.e., incorporating both the geomagnetic as well as an Askaryan-type \v{C}erenkov radiation mechanism. Initial results at energies around 10~TeV have been presented in \citep{KonstantinovArena2005}.

This article describes the next step (REAS2) in the development of our geosynchrotron radiation model. The well-tested radio emission part of the Monte Carlo code is essentially kept the same as in REAS1, but the analytical description of the air shower characteristics is replaced by detailed multi-dimensional histograms obtained from state-of-the-art CORSIKA \citep{HeckKnappCapdevielle1998} simulations. The much more realistic air shower model allows calculations of the radio emission with much higher precision. At the same time, the code allows us to make a very gradual transition from the earlier, well-understood and thoroughly tested simulations, providing insights in the importance of various aspects of particle distributions in EAS for the radio signal.

In this article, we first discuss the strategy of Monte Carlo simulations with REAS2 and then analyse in detail how the improvement of the underlying air shower model, the elimination of the previously needed track length parameter, and the adoption of the true electron to positron ratio change the predicted radio emission. In the second part we illustrate how the new simulations can be used to analyse different regimes identifiable in the particle distributions and their contribution to the radio pulses, before concluding with an outlook on the future development.

\section{CORSIKA-based Monte Carlo simulations}

With REAS2, the process of simulating radio emission from extensive air showers is separated into two steps. First, the air shower is simulated as usual with CORSIKA, adopting the desired interaction models and simulation parameters. A tailor-made interface code is used to collect the relevant particle information in histograms during the air shower simulation. At the end of the simulation, the histograms are saved to disk in a compact data file based on ROOT data structures.

The following particle information is collected for electrons and positrons separately in (usually) 50 slices equidistantly distributed in slant atmospheric depth over the air shower evolution (from first interaction to observer level):

\begin{itemize}
\item{one three-dimensional histogram of}
  \begin{enumerate}
  \item{particle arrival time relative to that of an imaginary primary particle propagating with the speed of light from the point of first interaction}
  \item{lateral distance of the particle from the shower core}
  \item{particle energy}
  \end{enumerate}
\item{and one three-dimensional histogram of}
  \begin{enumerate}
  \item{angle of the particle momentum to the shower axis}
  \item{angle of the particle momentum to the (radial) outward direction}
  \item{particle energy}
  \end{enumerate}
\end{itemize}

With the first set of histograms we have access to a four-dimensional histogram of atmospheric depth, particle arrival time, lateral distance and energy. In addition, we have information on the particle momentum angles to the shower axis and outward direction as a function of atmospheric depth and particle energy. For atmospheric depth values between the 50 slices, the histograms are interpolated appropriately. To improve the sampling of the shower evolution, the shower size longitudinal profile is taken from CORSIKA directly (sampled usually every 5~g~cm$^{-2}$).

Naturally, one complete higher-dimensional histogram would be superior to two separate lower-dimensional histograms. The approximation made by separating the information into the two above-mentioned histograms is, however, motivated by air shower physics and allows us to describe the air shower properties well with only moderate memory usage. Regarding the angular distribution of particle momenta, we chose to retain its correlation with particle energy rather than that with lateral distance. The correlation with particle energy is very strong as illustrated in \citep{NerlingBluemerEngel2006}, Fig.\ 11. An additional correlation with lateral distance certainly exists, but is neglected here. The radio signal is dominated by the bulk of the particles close to the shower axis, whereas particles at large lateral distances contribute only weakly (cf.\ Sec.\ \ref{rings}). For these particles close to the shower axis, however, the correlation with lateral distance can only be weak; otherwise, the angular distributions plotted in Fig.\ 11 of \citep{NerlingBluemerEngel2006} would show a stronger evolution with shower age. The current approximation gives a good representation of rotationally symmetric showers. Possible asymmetry effects due to the geomagnetic field (such as a systematic shift of electrons with respect to positrons) are not considered, but will be included at a later time. In any case, the histograms retain the systematic outward drift of particles from the shower core. %The histograms wash out azimuthal asymmetries in the particle distributions. At a later time, systematic shifts between the electron and positron distributions can be incorporated fairly easily. 

At energies of about $10^{16}$~eV and higher, the thinning option in CORSIKA is used to keep computation times acceptable. As we are only interested in the histogrammed particle distributions, the requirements on the quality of the thinned simulations is only moderate. Optimum $10^{-6}$ thinning \citep{Kobal2001} yields high-quality simulations clearly detailed enough for the calculation of histogrammed particle distributions and was therefore used in all simulations presented here.

In a second step, these histograms are imported in the REAS2 code, where they are used to recreate particles which follow the given distributions and then calculate the radio emission from the deflection of the particles in the earth's magnetic field. The well-understood and thoroughly tested radio emission calculation is essentially unchanged from our earlier code (for details see \citep{HuegeFalcke2005a}).

This two-step approach has a number of advantages. In particular, it allows us to make a very gradual transition from fully parametrised air showers to fully histogrammed air showers by switching on the histogrammed distributions individually one after the other and evaluating the changes arising in the radio signal. The fact that one does not have to re-run the complete air shower simulation each time one wants to calculate the radio emission in a different configuration (e.g., with parametrised vs.\ histogrammed distributions) is also very helpful in keeping the computation time low. A reduction of computation times has also been achieved by the implementation of a new time-grid algorithm and an algorithm to automatically optimise the time window and time resolution for a specific radio pulse calculation as a function of lateral observer distance from the shower centre. On a current standard PC (2 GHz class, 512 MB memory), the time distribution radio pulse calculation (without the CORSIKA run) takes approximately 36-48 hours for 200 observer positions (antennas).

\section{Effects of the improved distributions} \label{individual}

We investigate the effects arising due to the transition from parametrised distributions to CORSIKA-based histograms for the example of a typical $10^{17}$~eV vertical proton-induced air shower with shower maximum at $\sim 640$~g~cm$^{-2}$. A magnetic field of 0.48 Gauss with an inclination angle of 64.7$^{\circ}$, as appropriate for the LOPES-location in Karlsruhe, Germany, is adopted for all simulations.

All CORSIKA simulations were carried out with the QGSJETII.03 \citep{Ostapchenko2005,Ostapchenko2006b} high-energy and UrQMD1.3.1 \citep{Bass1998,Bleicher1999} low-energy interaction models. Electromagnetic cascades were treated with EGS4 \citep{EGS4} down to a threshold of 400~keV kinetic energy, where the particles are no longer relativistic and therefore become unimportant for the geosynchrotron mechanism.

To illustrate the changes arising in the transition from the parametrised to the histogrammed air shower model, we investigate the particle distributions individually, switching from parametrised to histogrammed distributions one after the other in a series of five steps. In the diagrams, we mark distributions which are parametrised with a ``0'' and distributions which are histogrammed with a ``1'' inside a 5-digit key. The order of the digits in the key corresponds to the order in which the histogrammed distributions replace the parametrised ones in the subsections \ref{sec:individual:start} to \ref{sec:individual:end}.
 
 In each of the five steps we look at the unlimited-bandwidth radio pulses at a close distance of 75~m and at a larger distance of 525~m to the north from the shower centre. The pulses at 525~m distance are amplified by a factor of 20 for a clearer illustration. These two distances represent different regimes for experimental measurements. Closer distances are particularly important for LOPES, while larger distances are relevant for radio detection on large scales as currently planned for the Pierre Auger Observatory.
 
 In a sixth step, we then investigate how the transition to the histogrammed particle distributions changes the pulses seen by observers 75~m and 525~m to the west from the shower centre.

\subsection{Air shower longitudinal evolution} \label{sec:individual:start}

The air shower longitudinal development in REAS1 was parametrised with a Greisen function \citep{HuegeFalcke2003a,Greisen1956} which was originally designed to describe purely electromagnetic showers. The position of the shower maximum in REAS1 was a user-defined parameter. Using this parameter one could mimic the effect of different primary particle masses, yet only in a crude way. Shower-to-shower fluctuations were not taken into account.

Air showers simulated with CORSIKA naturally take into account differences in the shower evolution related to the primary particle mass. Also, shower-to-shower fluctuations and the influence of different high- and low-energy interaction models are automatically accounted for. For the analyses presented in this article, we simulated a number of vertical $10^{17}$~eV proton-induced CORSIKA showers and then selected one with a typical profile and a shower maximum close to the expected average depth of $\sim 640$~g~cm$^{-2}$.

\begin{figure}[htb]
\begin{minipage}{15.5pc}
\includegraphics[angle=270,width=15.5pc]{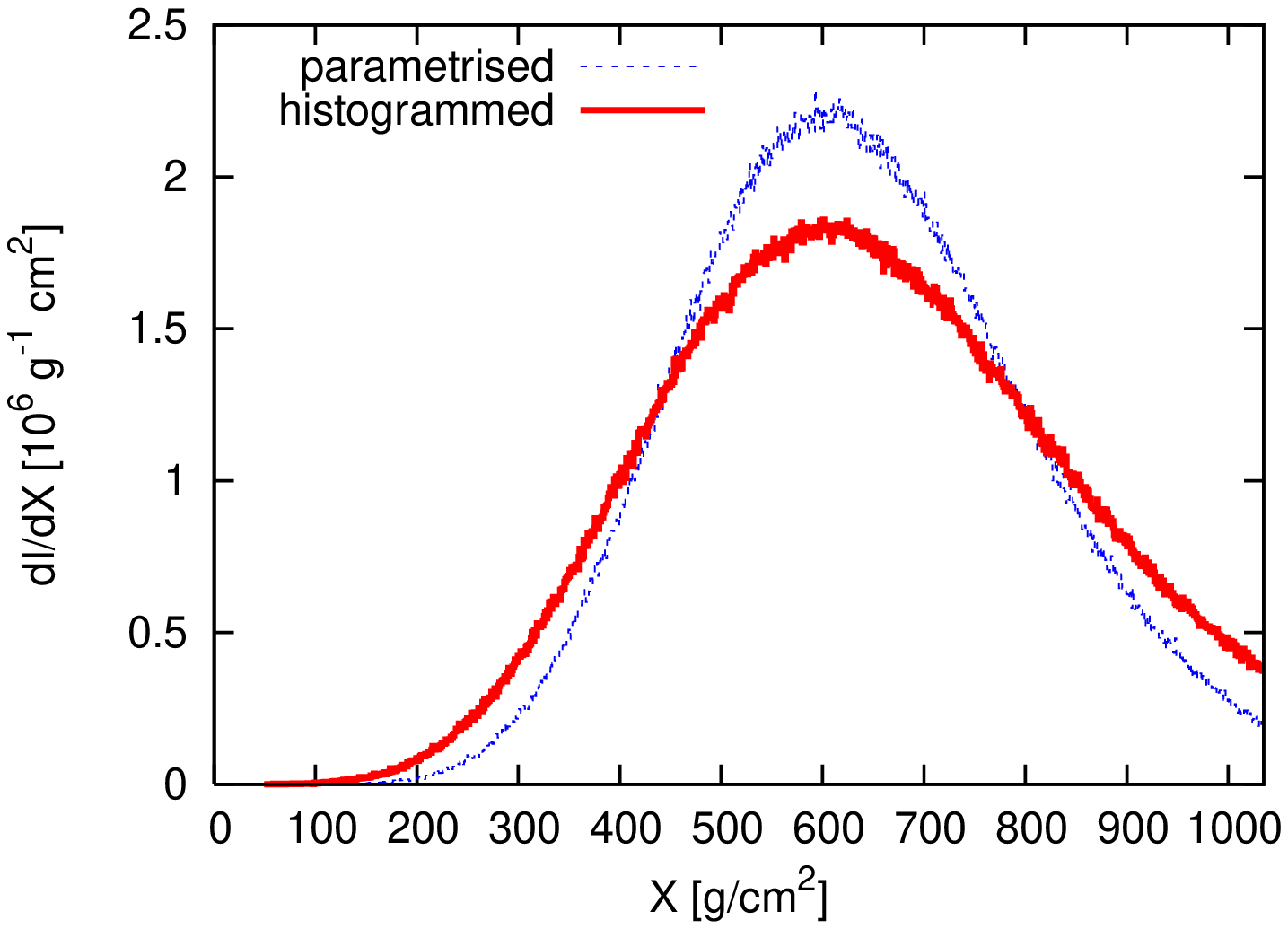}
\caption{\label{injechisto}Particle (sum of e$^{+}$ and e$^{-}$) injection profiles in parametrised and histogrammed versions.}
\end{minipage} \hspace{1.5pc}
\begin{minipage}{15.5pc}
\includegraphics[angle=270,width=15.5pc]{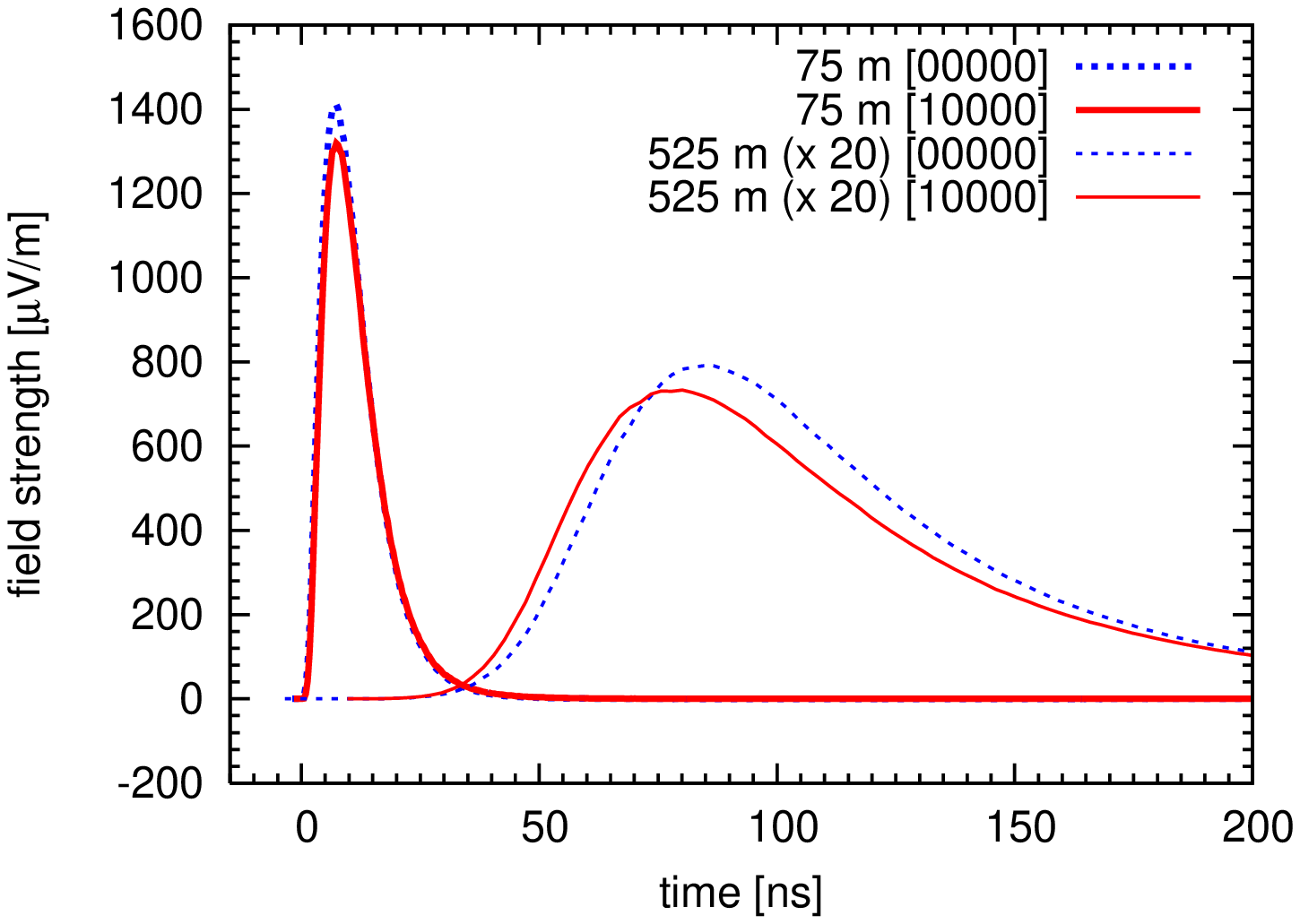}
\caption{\label{injecpulses}Radio pulses for parametrised and histogrammed air shower evolutions.}
\end{minipage}
\end{figure}

In figure \ref{injechisto} we compare the parametrised and CORSIKA-based particle injection profiles, which are closely related to the corresponding air shower longitudinal profiles via the track-length parameter $\lambda$ (cf.\ Sec.\ \ref{elemtrack}; throughout section \ref{individual}, $\lambda$ is fixed to 36.7~g~cm$^{-2}$ with an exponential distribution of particle track lengths). The parametrised profile and the profile of the selected CORSIKA shower have virtually identical integrated particle numbers, but the CORSIKA-based profile is slightly wider with a lower particle number at shower maximum.

Figure \ref{injecpulses} shows the radio pulses\footnote{Throughout this article we show radio pulses corresponding to the east-west polarisation component of the emission. This allows us to retain the sign of the electric field while its strength is virtually identical to the total field strength (in case of vertical showers). Positive field strengths denote electric fields pointing to the west. The zero-point in time is arbitrarily set to the start of the 75~m pulses.} corresponding to a fully parametrised shower (00000) and a shower with histogrammed air shower evolution profile, but all other distributions parametrised (10000). The changes to the pulse shapes are very small.

\subsection{Particle energies}

We now additionally switch the particle energy distribution from parametrised to histogrammed. Please note that the histogrammed energy distribution changes as a function of atmospheric depth, while the parametrised distribution was used globally throughout the shower.

\begin{figure}[htb]
\begin{minipage}{15.5pc}
\vspace{0.6pc}
\includegraphics[width=15.5pc]{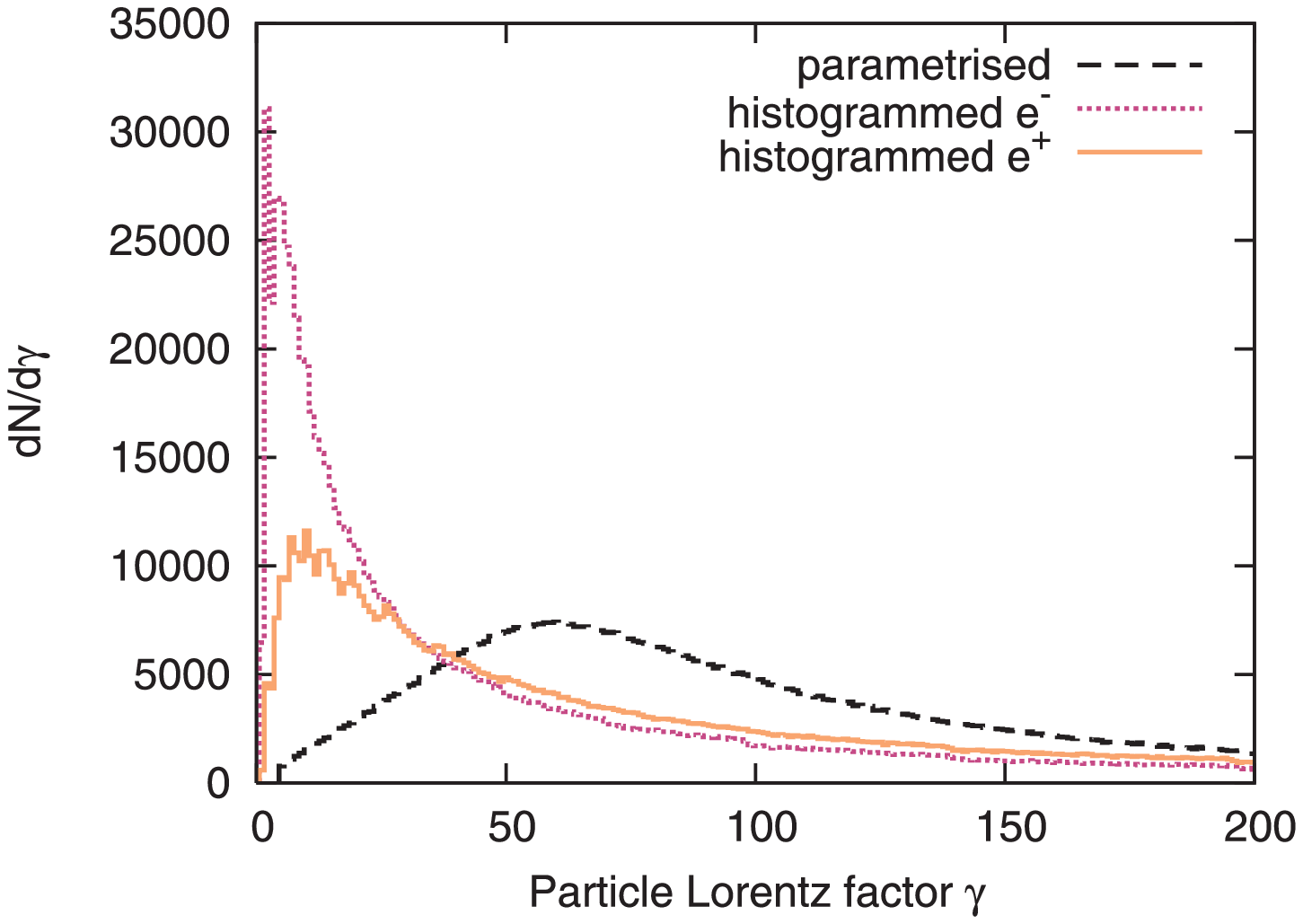}
\caption{\label{energyhisto}Distribution of particle energies in parametrised and histogrammed versions (averaged over all other distributions, normalised to a total of $10^{6}$ particles).}
\end{minipage} \hspace{1.5pc}
\begin{minipage}{15.5pc}
\includegraphics[angle=270,width=15.5pc]{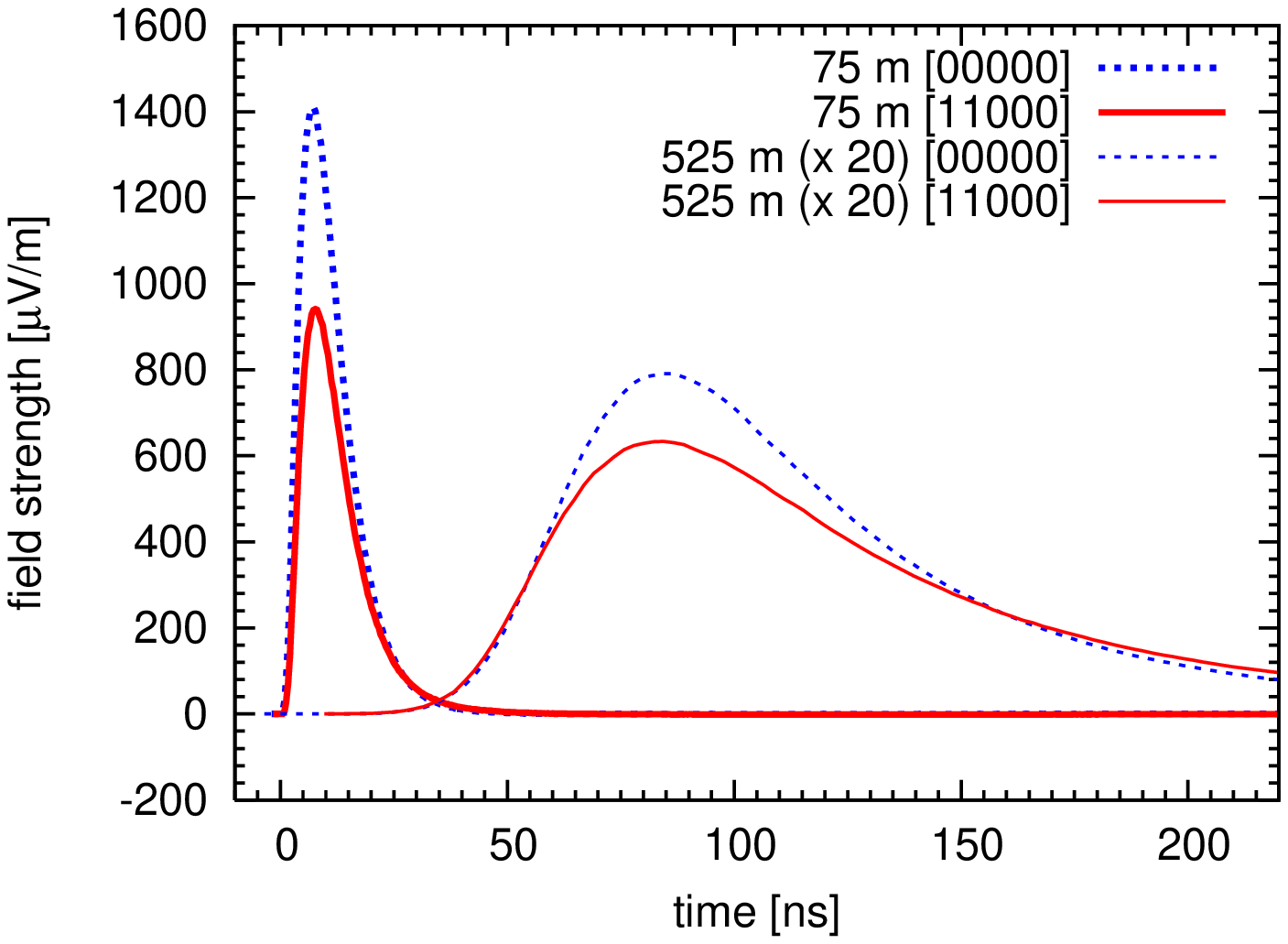}
\caption{\label{energypulses}Radio pulses for fully parametrised showers compared with those for showers with histogrammed shower evolution and energy distribution.}
\end{minipage}
\end{figure}

In figure \ref{energyhisto} we compare the parametrised particle energy distribution with the histogrammed particle energy distributions as obtained from CORSIKA. The histogrammed energy distributions deviate significantly from the parametrisation. In particular, there are more particles at lower energies and there is no peak in the distribution at medium Lorentz factors.

Figure \ref{energypulses} shows the effect of switching on the histogrammed energy distribution. The changes are only slight. The relative up-weighting of low energies, and, correspondingly, down-weighting of higher energies is mainly visible in a reduced pulse height in the shower centre, where high-energy particles with narrow beaming cones contribute strongly (cf.\ Sec.\ \ref{pulseshapeanalysis}). The disappearance of the prominent peak in the energy distribution around Lorentz factors of 60 is important for the pulse shapes along the east-west axis from the shower centre (cf.\ Sec.\ \ref{pulseshapes}).

\subsection{Particle lateral distribution}

In this step, we additionally switch on the histogrammed lateral particle distribution. The parametrised distribution already accounted for changes with atmospheric depth through the variation of the Moli\`ere radius. The histogrammed distribution takes into account both the development with atmospheric depth and the correlation with the particle energy distribution.

\begin{figure}[htb]
\begin{minipage}{15.5pc}
\vspace{0.6pc}
\includegraphics[width=15.5pc]{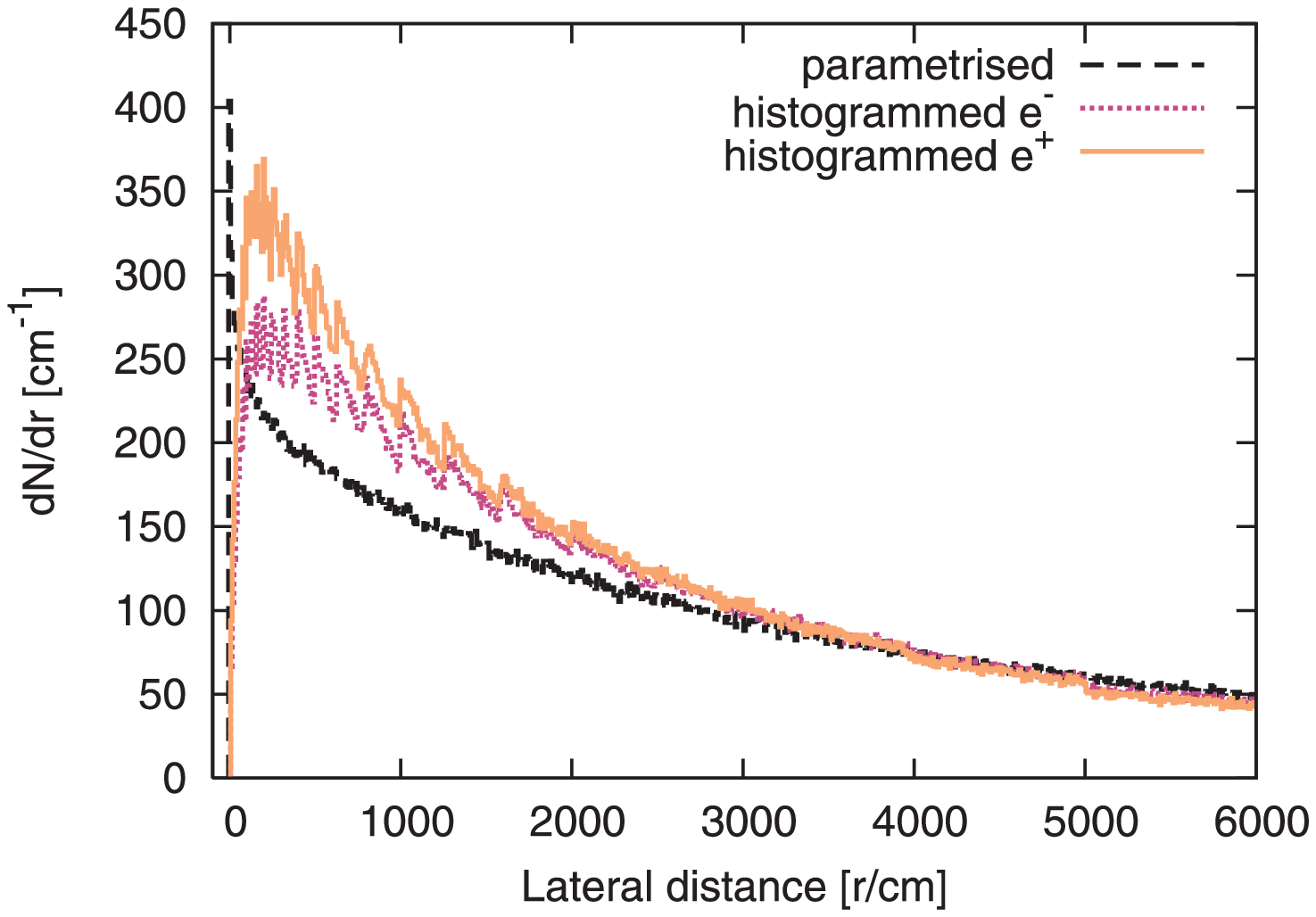}
\caption{\label{lateralhisto}Lateral particle distribution in parametrised and histogrammed versions (averaged over all other distributions, normalised to a total of $10^{6}$ particles).}
\end{minipage}\hspace{1.5pc}
\begin{minipage}{15.5pc}
\includegraphics[angle=270,width=15.5pc]{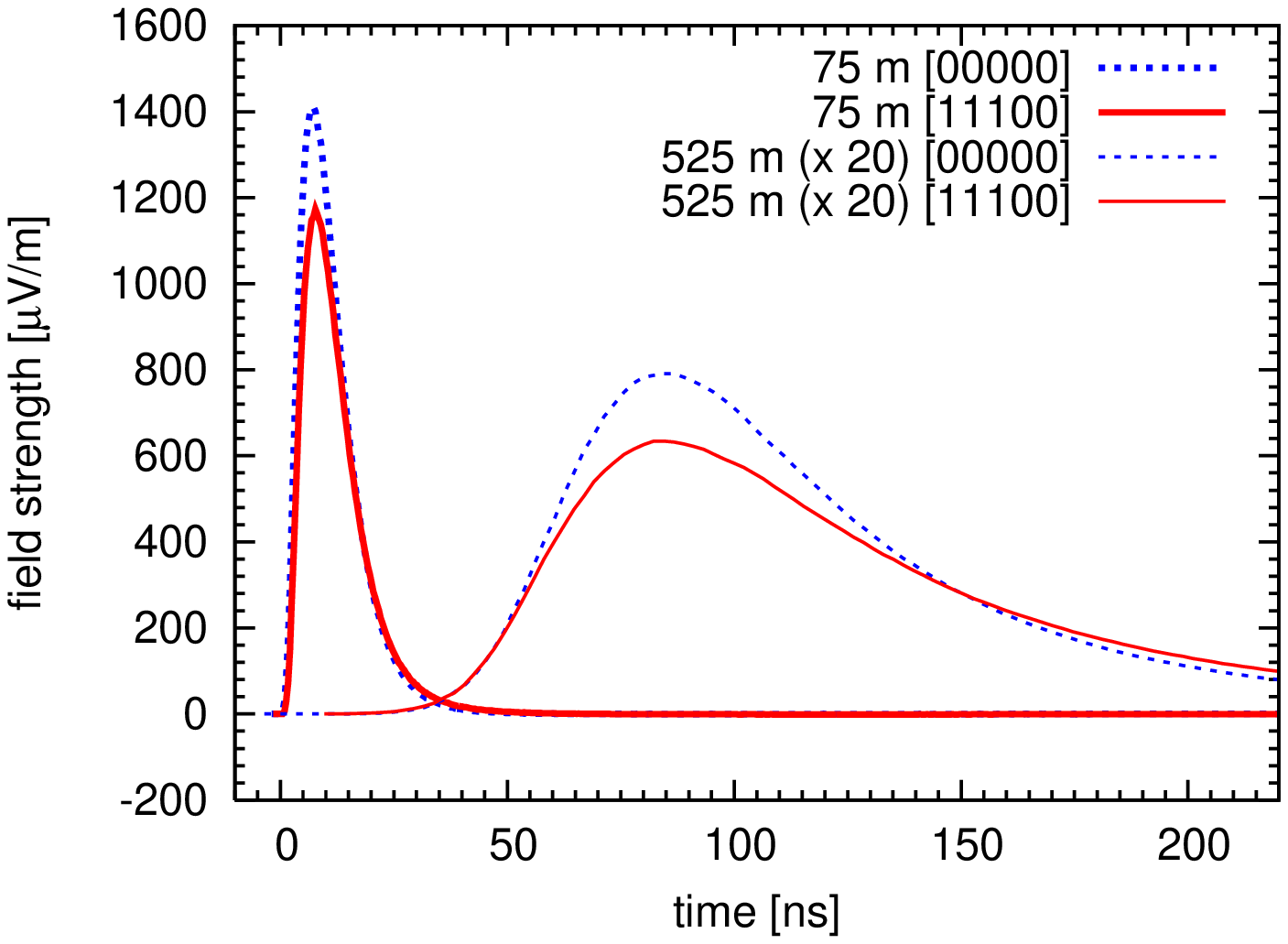}
\caption{\label{lateralpulses}Radio pulses for fully parametrised showers compared with those for showers with histogrammed shower evolution, energy and lateral distributions.}
\end{minipage}
\end{figure}

In figure \ref{lateralhisto} we compare the lateral distributions of shower particles in the parametrised and histogrammed versions.\footnote{The ``zig-zag'' structure present in the histogrammed distributions shown throughout this article is an artifact of the adopted logarithmic histogram binning. As the logarithmic bin boundaries have no regular structure on a linear scale, this is unproblematic for the calculation of the radio signal. In most cases logarithmic binning gives a better representation of the particle distributions than using linear intervals.} The parametrisation used is an NKG-function as given in \citep{Greisen1956,KamataNishimura1958}. The histogrammed distributions have a higher particle density in the medium-distance region, but a lower density in the extreme centre.

Figure \ref{lateralpulses} shows the corresponding radio pulses. The pulse heights in the shower centre are slightly amplified with respect to Fig.\ \ref{energypulses}, because the histogrammed lateral distribution is characterised by more high-energy particles near the shower core.

\subsection{Particle arrival times}

We now additionally switch the particle arrival time distribution from parametrised to histogrammed. The parametrised distribution was only correlated with radial distance from the shower axis. In particular, it was assumed to be the same over the full shower evolution. In contrast, the histogrammed arrival time distribution is now fully correlated with atmospheric depth, lateral distance and particle energy.

\begin{figure}[htb]
\begin{minipage}{15.5pc}
\vspace{0.6pc}
\includegraphics[width=15.5pc]{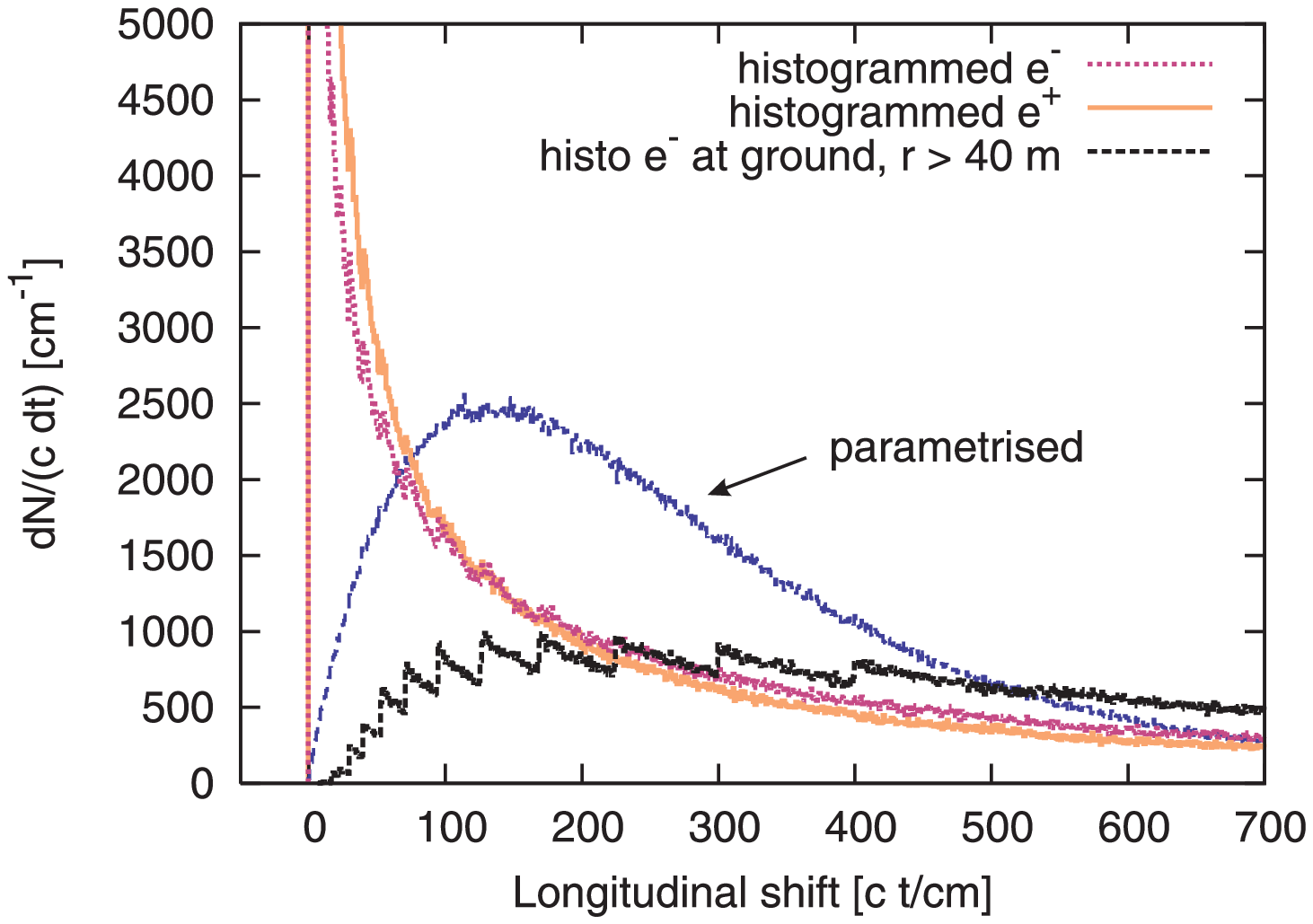}
\caption{\label{timehisto}Distribution of particle arrival times in parametrised and histogrammed versions (averaged over all other distributions if not stated otherwise, normalised to a total of $10^{6}$ particles).}
\end{minipage} \hspace{1.5pc}
\begin{minipage}{15.5pc}
\includegraphics[angle=270,width=15.5pc]{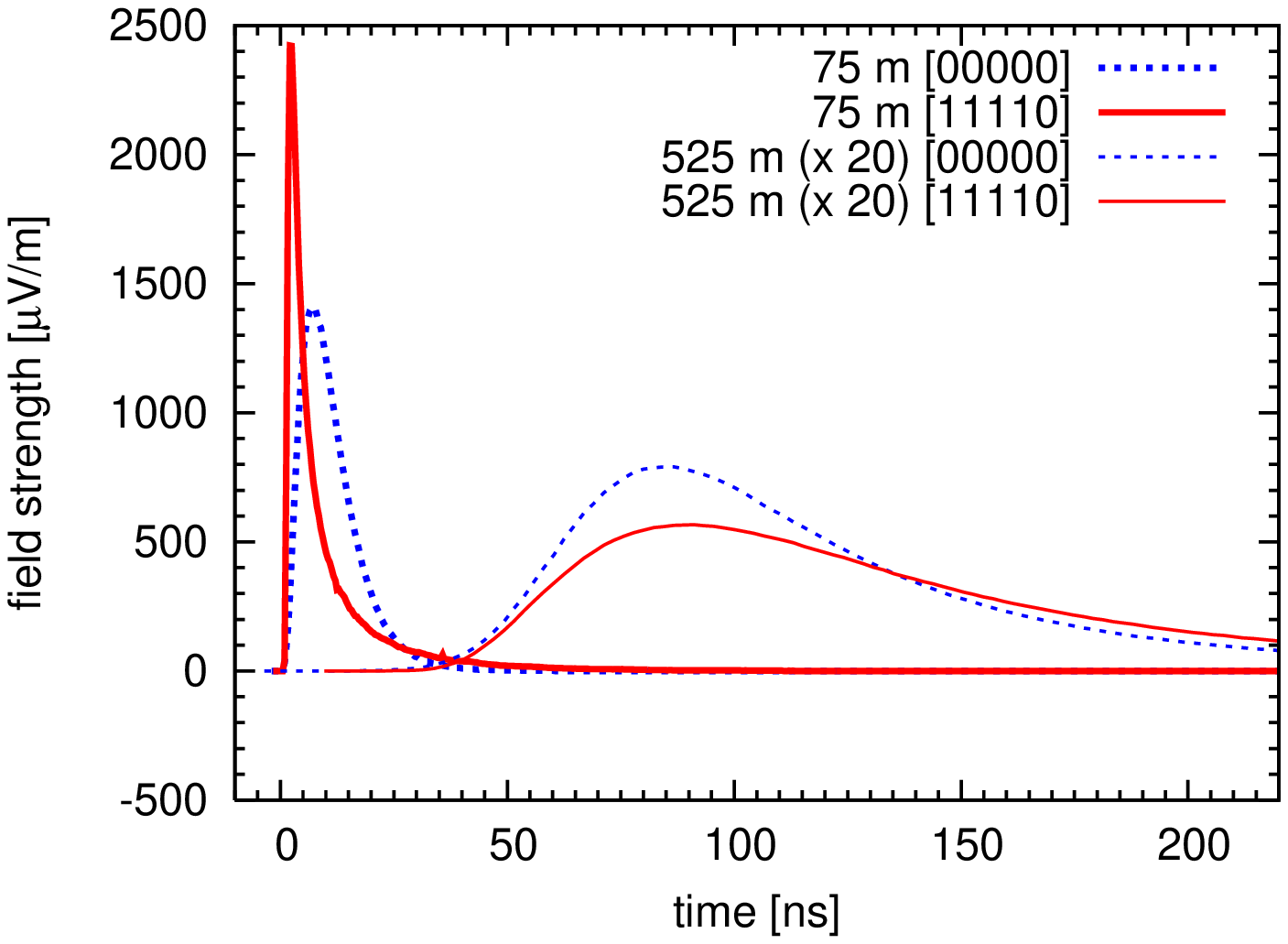}
\caption{\label{timepulses}Radio pulses for fully parametrised showers compared with those for showers with histogrammed shower evolution, energy, lateral and arrival time distributions.\newline}
\end{minipage}
\end{figure}

The arrival time distribution has strong influence on the shape (and thus frequency spectrum) of the radio pulses (for a discussion, see \citep{HuegeFalcke2003a}). Figure \ref{timehisto} shows the parametrisation of the particle arrival times used so far in comparison with the distributions obtained from CORSIKA. The CORSIKA-based overall arrival time distributions are much narrower. This was to be expected because the parametrised distribution \citep{HuegeFalcke2003a} was derived from measurements \citep{AgnettaAmbrosioAramo1997} that are very difficult if not impossible to perform close to the shower axis because of problems associated with detector saturation and the definition of the time reference given by the arrival of the leading particle. The measurements available in the literature therefore represent the arrival time distributions at somewhat larger distances, where the distributions are significantly broader. In addition, these measurements registered all charged particles, including muons. At increasing distances from the shower core, the muons arrive systematically earlier than electrons and positrons and thus additionally widen the arrival time distribution. In agreement with our expectations, the CORSIKA-derived arrival time distribution for electrons (and positrons) with distances greater than 40~m from the shower axis becomes qualitatively similar to the parametrisation (and thus measurements).
 
The differences in particle arrival time directly affect the shape of the radio pulses as plotted in Fig.\ \ref{timepulses}. In particular, close to the shower centre, where geometrical time delays are of no importance, the pulse shape directly reflects the particle arrival time distribution and thus becomes much narrower with a steeper leading edge in the histogrammed version. (Atmospheric refraction which is expected to slightly smear out the pulses is not taken into account here.)

Additionally, the pulses amplitudes close to the shower centre are increased significantly with respect to those shown in Fig.\ \ref{lateralpulses}. This is a consequence of the strong correlation between the energy, lateral and arrival time distributions, the combination of which preferably puts high-energy particles with little time lag close to the shower core. These particles then contribute strongly to the radio pulses close to the shower centre. If only one of the three distributions is switched back to parametrised, this amplification effect diminishes, illustrating how important a realistic representation of the strongly correlated particle distributions is.

\subsection{Particle momentum angles to the shower axis} \label{sec:individual:end}

Finally, we additionally switch on the histogrammed particle momentum angle distribution. In the parametrised version, the  particle momentum direction (at the start of a particle trajectory) was directly linked to the position of particle creation. The histogrammed momentum angle distribution, on the other hand, is strongly correlated with the particle energy distribution.

\begin{figure}[htb]
\begin{minipage}{15.5pc}
\vspace{0.6pc}
\includegraphics[width=15.5pc]{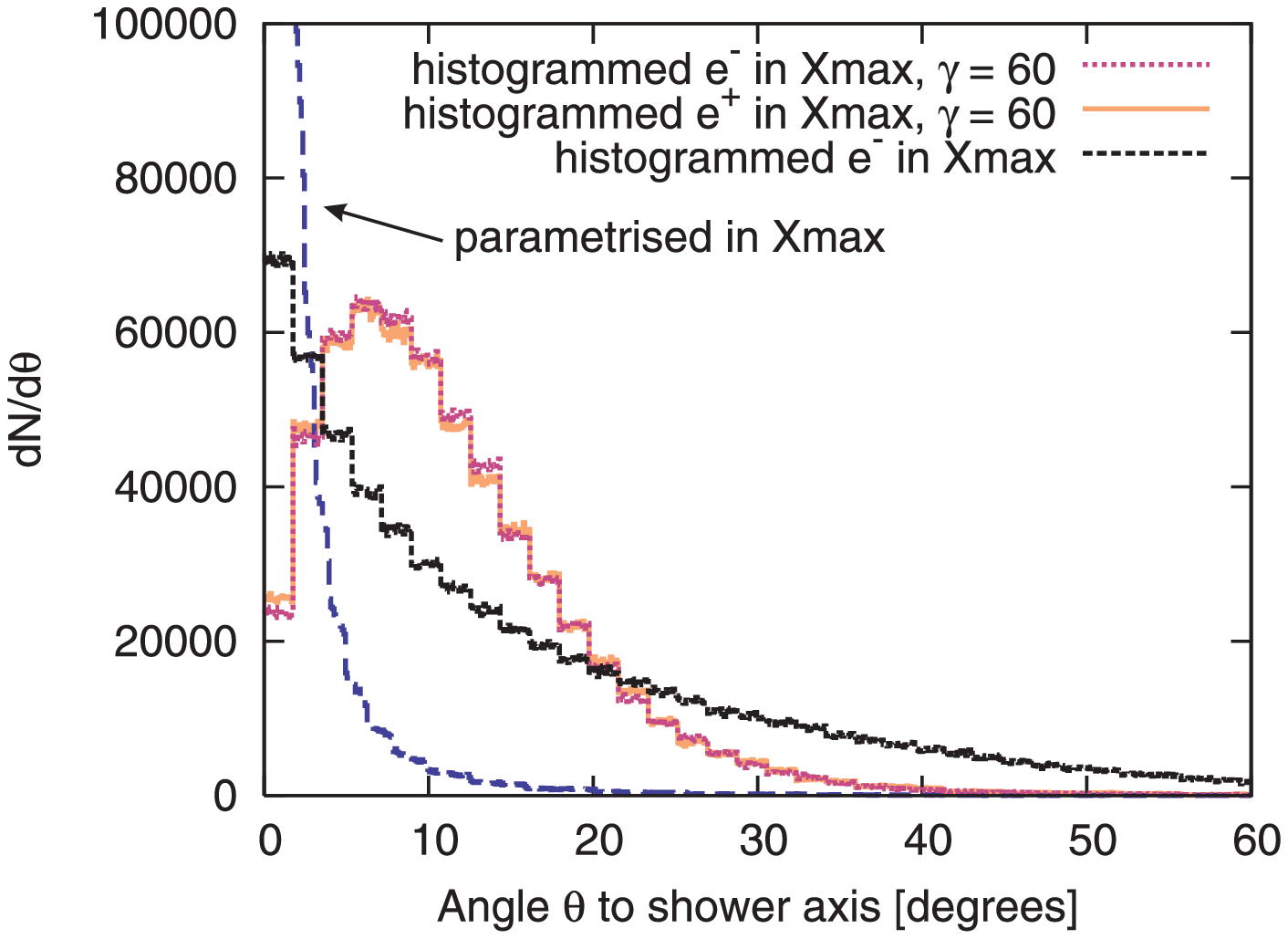}
\caption{\label{anglehisto}Particle momentum angles to the shower axis at particle creation (averaged over all distributions except atmospheric depth if not stated otherwise, normalised to a total of $10^{6}$ particles).}
\end{minipage} \hspace{1.5pc}
\begin{minipage}{15.5pc}
\includegraphics[angle=270,width=15.5pc]{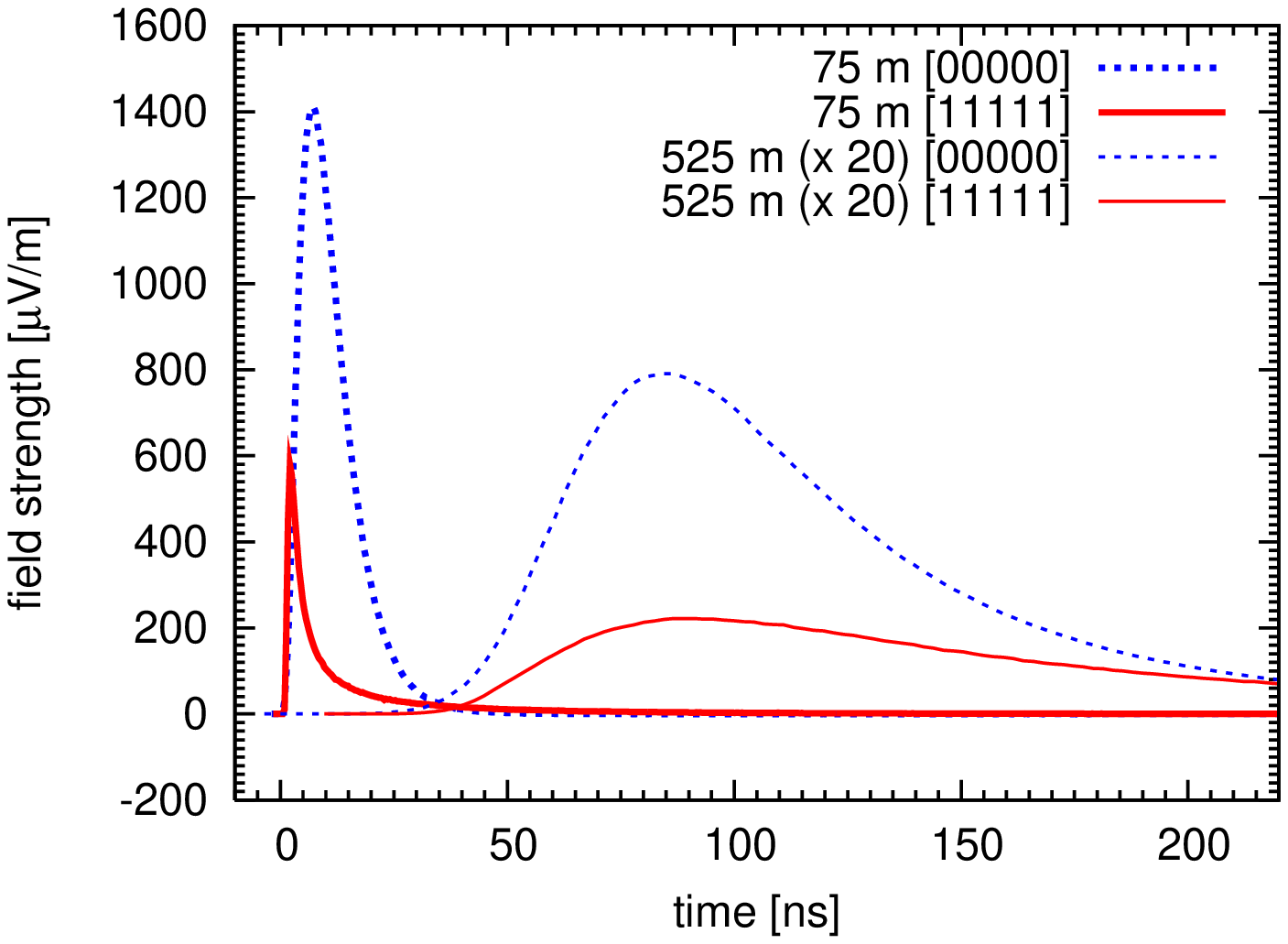}
\caption{\label{anglepulses}Radio pulses for fully parametrised showers compared with those for fully histogrammed showers.\newline\newline\newline}
\end{minipage}
\end{figure}

Figure \ref{anglehisto} shows the angular distribution of particle momenta at particle creation relative to the shower axis. In the parametrised version, the particle momenta are very collimated. The histogrammed distribution is much wider. Its strong correlation with particle energy is illustrated by a comparison of the overall distribution of electrons with that of particles with a Lorentz factor of 60.

Correspondingly, and expectedly, the effect on the radio pulses is strong, as depicted in Fig.\ \ref{anglepulses} in comparison with Fig.\ \ref{timepulses}. The emission of relativistic particles is beamed into a narrow cone along the particle momentum direction. In the histogrammed case, the momentum direction is much less aligned with the shower axis, and therefore a significant fraction of the radio signal is emitted at large angles to the shower axis. As a consequence, the pulses are damped significantly, especially in the centre region, where the high-energy particles with especially narrow beaming cones contribute most.

\subsection{Changes along the east-west axis} \label{pulseshapes}

So far we have only looked at pulses to the north of the shower centre. Here (and likewise to the south), switching from the parametrised to the histogrammed particle distributions mainly changes the pulse amplitudes, but leaves the pulse shapes similar. The situation can be different to the east or west of the shower centre. There, the contributions from electrons and positrons interact in a much more complex way: the geometry under which the observer sees electrons and positrons, which themselves bend to the east and west in the geomagnetic field, becomes asymmetric. (See \citep{HuegeFalcke2005a} for a detailed discussion.)

In simulations with REAS1, pulses at moderate to large distances in the east or west from the shower centre could become bipolar, depending on the specific air shower configuration. The reason is that the pulses emitted by individual particles have a slightly bipolar structure and, if an observer starts to see only part of these bipolar pulses, the polarity of the overall pulse arising from the superposition of many individual particle pulses can change. (An example is shown in Fig.\ 18 of \citep{HuegeFalcke2005a} together with a detailed discussion.)

Such effects, however, can only occur if the air shower model (more precisely its particle distributions) exhibit clear ``structures''. This is the case for the parametrised distributions: the angular distribution of particle momenta is very narrow and the particle energy distribution has a prominent peak around Lorentz factors of 60. Even more pronounced bipolarities can be produced when the air shower model is simplified further, e.g., by neglecting the lateral structure of the air shower or running the simulation with long, fixed particle track lengths. As the CORSIKA-derived histogrammed particle distributions do not show any such prominent ``structures'', the bipolar pulses do not occur in simulations of CORSIKA-histogram based showers.

\begin{figure}[htb]
\begin{minipage}{15.5pc}
\includegraphics[angle=270,width=15.5pc]{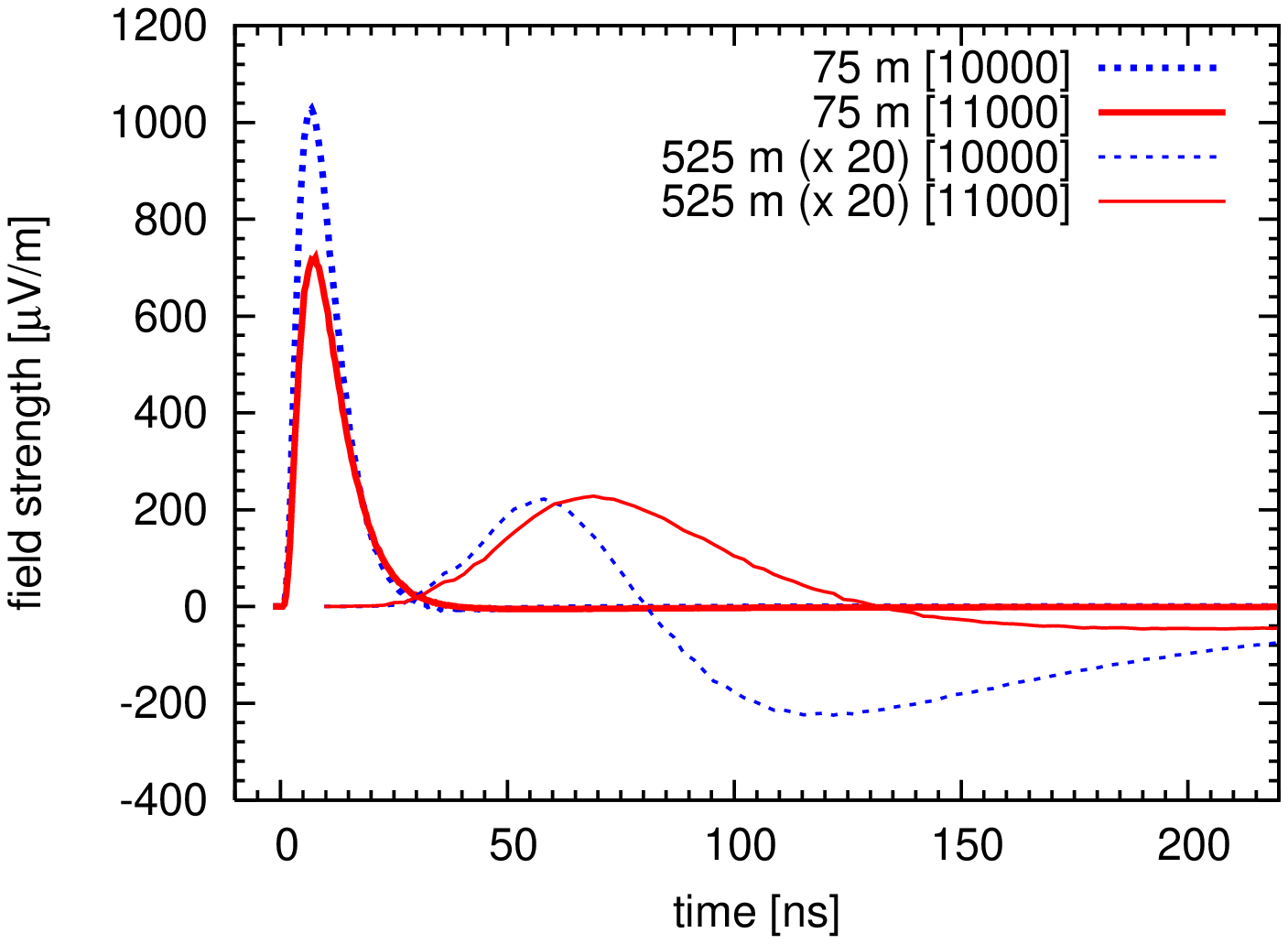}
\caption{\label{energypulses90deg}Influence of the histogrammed energy distribution on the radio pulses to the west of the shower centre.\newline}
\end{minipage} \hspace{1.5pc}
\begin{minipage}{15.5pc}
\includegraphics[angle=270,width=15.5pc]{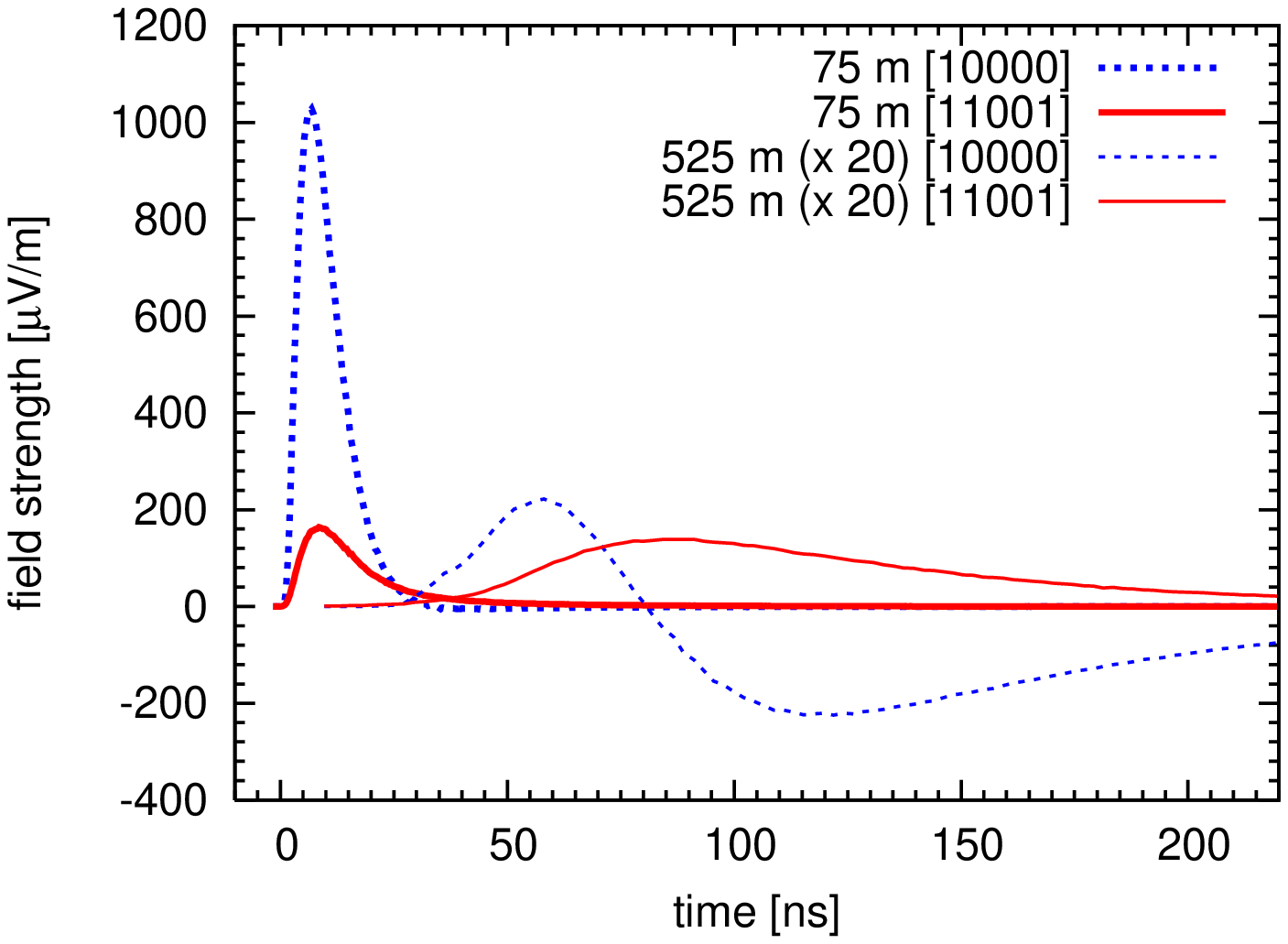}
\caption{\label{anglepulses90deg}Influence of the histogrammed energy and momentum angle distributions on the radio pulses to the west of the shower centre.}
\end{minipage}
\end{figure}

Figure \ref{energypulses90deg} shows the changes of pulses to the west from the shower centre associated with switching from the parametrised to the histogrammed energy distribution. The bipolarity of the 525~m pulses is washed out significantly because the prominent peak in the energy distribution (Fig.\ \ref{energyhisto}) is no longer present. Similarly, figure \ref{anglepulses90deg} demonstrates the effect of additionally switching on the realistic momentum angular distribution. The bipolarity of the 525~m pulses has diminished even more. When all of the histogrammed distributions are switched on simultaneously, the bipolarity in the pulses vanishes completely. Bipolarities present in the REAS1 simulations can therefore be explained as an artifact of the simplified air shower model.

Having illustrated in detail the effects arising from the transition to realistic, histogrammed particle distributions, we now investigate a second major factor changing the radio pulses in the transition from REAS1 to REAS2.

\section{Eliminating the track length parameter} \label{elemtrack}

In simulations with REAS1, a free parameter, the so-called ``track length parameter'' $\lambda$, had to be set by the user. $\lambda$ denotes the average length (in g~cm$^{-2}$) of an individual particle trajectory in the simulation. It relates the longitudinal shower profile $N(X)$ (number of particles traversing an imaginary detector plane as a function of atmospheric depth $X$) as taken from the CORSIKA simulations to the longitudinal particle injection profile $I(X)$ (number of particles being created per d$X$ depth interval along the shower axis as a function of atmospheric depth $X$) via
\begin{equation} \label{eqn:inject}
\frac{\mathrm{d}N(X)}{\mathrm{d}X} = I(X) - \frac{N(X)}{\lambda},
\end{equation}
which ensures that the total track length integrated over all particles is constant irrespective of the value of $\lambda$. A large value leads to a simulation with few long tracks, whereas a small value corresponds to a simulation with many short tracks. Even though the normalisation is correct irrespective of the value of $\lambda$, the simulated radio signals can be influenced by the choice of a specific value. In particular, the angular distribution of particles evolves during their propagation along the trajectories. Consequently, the choice of a specific $\lambda$ indirectly influences the effective angular spread of the particle momenta. (For details on the role of the track-length parameter and the differences between fixed and exponentially distributed track lengths, see the discussion in \citep{HuegeFalcke2005a}.)

In REAS1 simulations, we usually set $\lambda$ to a value of 36.7~g~cm$^{-2}$ with an exponential distribution of particle track lengths. As explained above, this implicitly sets a characteristic angular spread for the particle momenta. It is, however not straight-forward to motivate the choice of a particular value for $\lambda$ either from experimental data or from simulations.

In REAS2 simulations, the situation is different. The CORSIKA-derived histograms contain detailed information on the angular spread of the particle momenta during the course of the full shower evolution. If $\lambda$ is set to inappropriately large values in this case, the angular spread of particle momenta (and also the distribution of particle energies) starts to deviate systematically from the histogrammed distributions in the course of the particle trajectories. The simulation can then become inconsistent.

\begin{figure}[htb]
\centering
\includegraphics[angle=270,width=18pc]{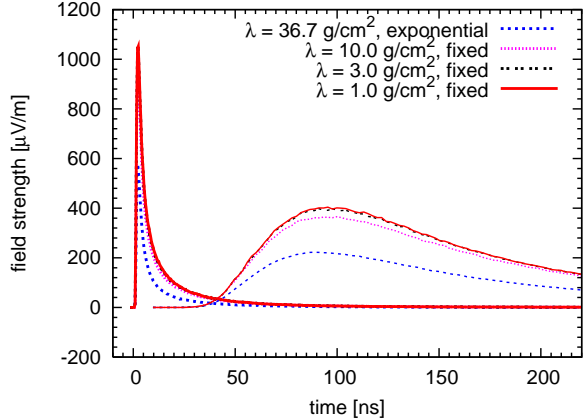}
\caption{\label{tracklengthpulses}Radio signal convergence in a fully histogrammed air shower when going from 36.7~g~cm$^{-2}$, exponentially distributed, to 1.0~g~cm$^{-2}$, fixed track lengths.}
\end{figure}

When $\lambda$ is set to a small value, on the other hand, each particle always follows the local particle distributions well throughout its full trajectory. Trajectories longer than $\lambda$ are then effectively described by an integration over multiple, independent representative segments of length $\lambda$. In REAS2, setting $\lambda$ to a sufficiently small value solves the problem of a particular choice for the $\lambda$ parameter in a natural way. The radio pulses converge and become independent of the value of $\lambda$ when its value is chosen smaller and smaller (cf.\ Fig.\ \ref{tracklengthpulses}). For air showers in the US standard atmosphere a value of 1~g~cm$^{-2}$ proves to give good results while the calculation is still numerically stable and efficient.

\section{Particle pairing}

In REAS1, electrons and positrons are always created in pairs --- a reasonable approximation of the situation in air showers. As explained in \citep{HuegeFalcke2003a}, part of the emission from electrons and positrons adds up, whereas other contributions cancel out. Pairwise creation of particles increases the numerical stability of calculating these cancellations. It is, however, well-known that air showers in fact exhibit a considerable negative charge excess of order $\sim 20$\% at low electron energy, which could influence the radio pulse strengths and polarisation properties.

The histograms obtained from CORSIKA contain separate information for electrons and positrons, including the differences in particle multiplicity. In REAS2, differences in the electron and positron distributions are therefore taken into account in detail. To assess the importance of these differences, the radio emission can be calculated in REAS2 with different strategies: electrons and positrons can always be created in pairs (``always paired'', as in REAS1), electrons and positrons can be created in pairs (when possible), but with the correct ratio of electrons to positrons in each bin of phase-space (``ratio paired''), and electrons and positrons can be created completely independently (``never paired'') with the correct ratio in each bin of phase-space.

\begin{figure}[htb]
\begin{minipage}{15.5pc}
\includegraphics[angle=270,width=15.5pc]{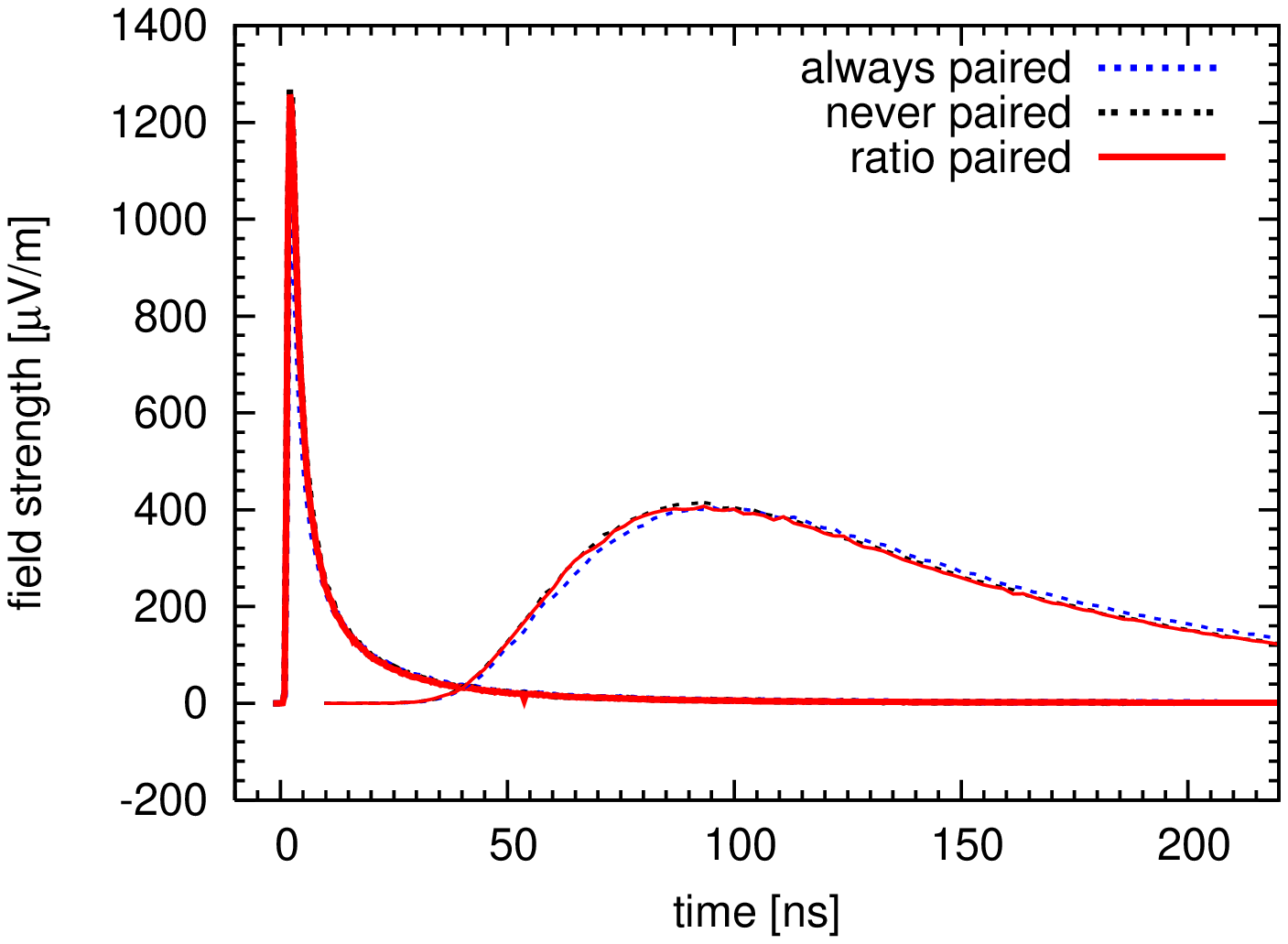}
\caption{\label{pairing}Differences between particle pairing modes in the north from the shower centre.}
\end{minipage} \hspace{1.5pc}
\begin{minipage}{15.5pc}
\includegraphics[angle=270,width=15.5pc]{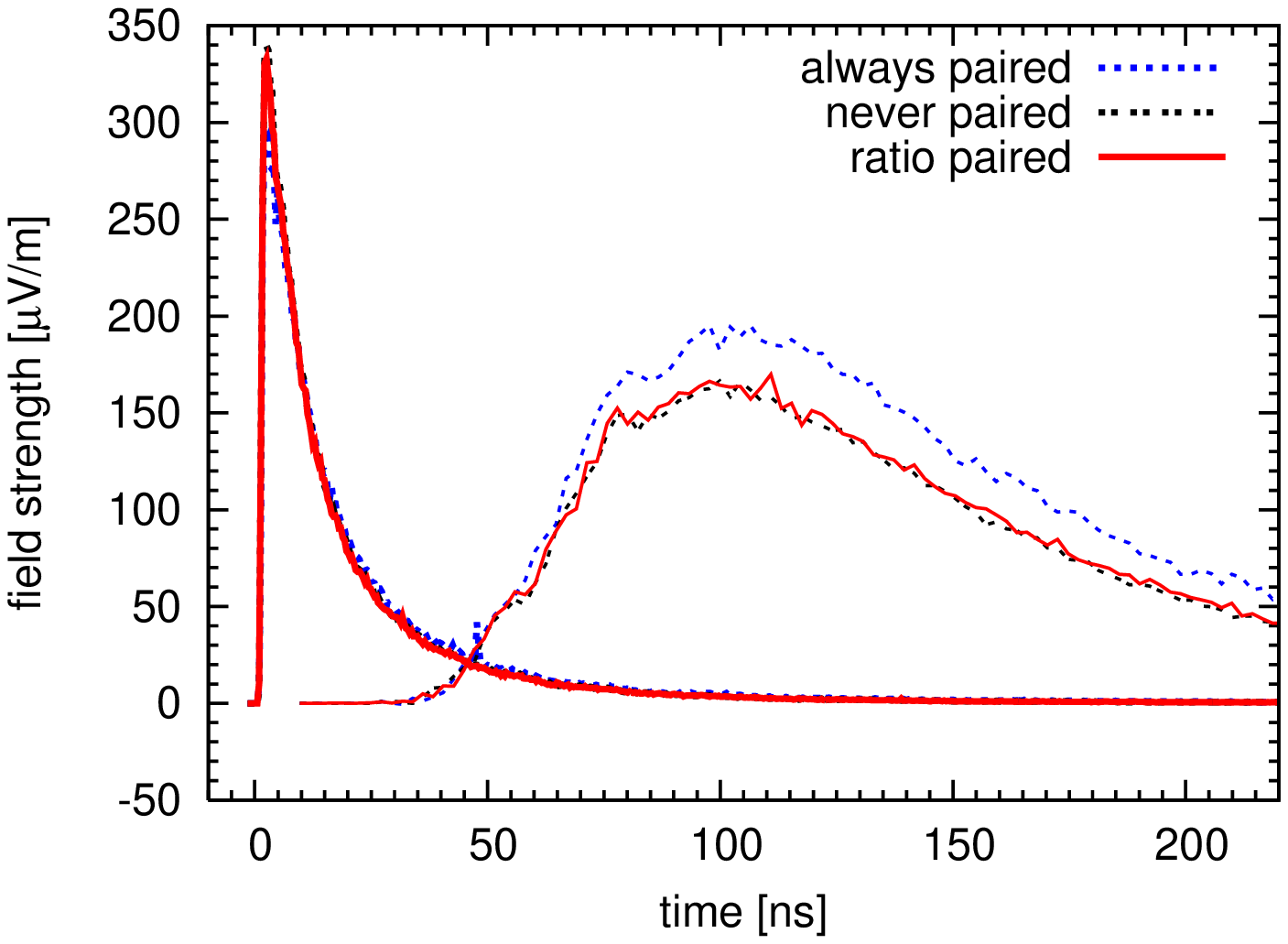}
\caption{\label{pairing90deg}Differences between particle pairing modes in the west from the shower centre.}
\end{minipage}
\end{figure}

Figure \ref{pairing} demonstrates the effect of the different ``pairing modes'' for observers to the north from the shower centre. The differences are negligible. For an observer to the west from the shower centre, the electron to positron ratio can be expected to have more influence because the geometry is asymmetric for the two species. A systematic difference is indeed visible between the ``always paired'' mode and the modes that correctly take into account the electron-to-positron ratio. Similar effects can be seen in the north-south polarisation component (not shown here), i.e., there is a slight effect on the polarisation properties of the emission. No significant difference is visible between the ``never paired'' and ``ratio paired'' modes, providing additional support for the validity of the ``small $\lambda$'' simulation strategy.

\section{Overall result}

So far we have only looked at isolated aspects of the air shower model and their influence on the calculated radio signal. Now we compare the overall changes between the radio emission from a fully parametrised shower (as in REAS1, with $\lambda = 36.7$~g~cm$^{-2}$, exponential track length distribution and ``always paired'' particles, see \citep{HuegeFalcke2005a,HuegeFalcke2005b}) and a fully histogrammed air shower (REAS2, with $\lambda = 1$~g~cm$^{-2}$ fixed track lengths and ``ratio paired'' particles).

\begin{figure}[htb]
\begin{minipage}{15.5pc}
\includegraphics[angle=270,width=15.5pc]{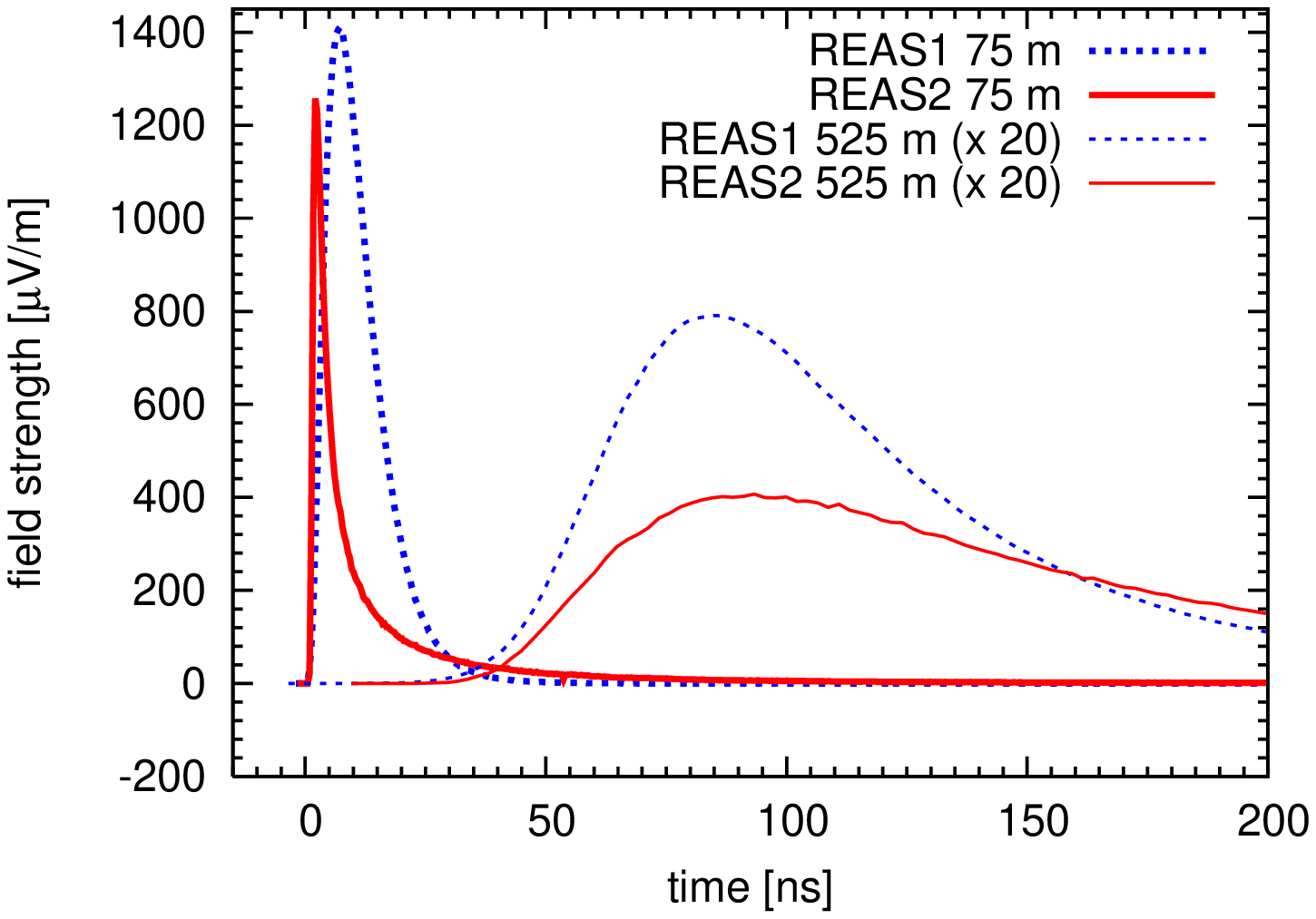}
\caption{\label{overallpulses}Overall change in the simulated radio pulses north of the shower centre between REAS1 and REAS2.}
\end{minipage} \hspace{1.5pc}
\begin{minipage}{15.5pc}
\includegraphics[angle=270,width=15.5pc]{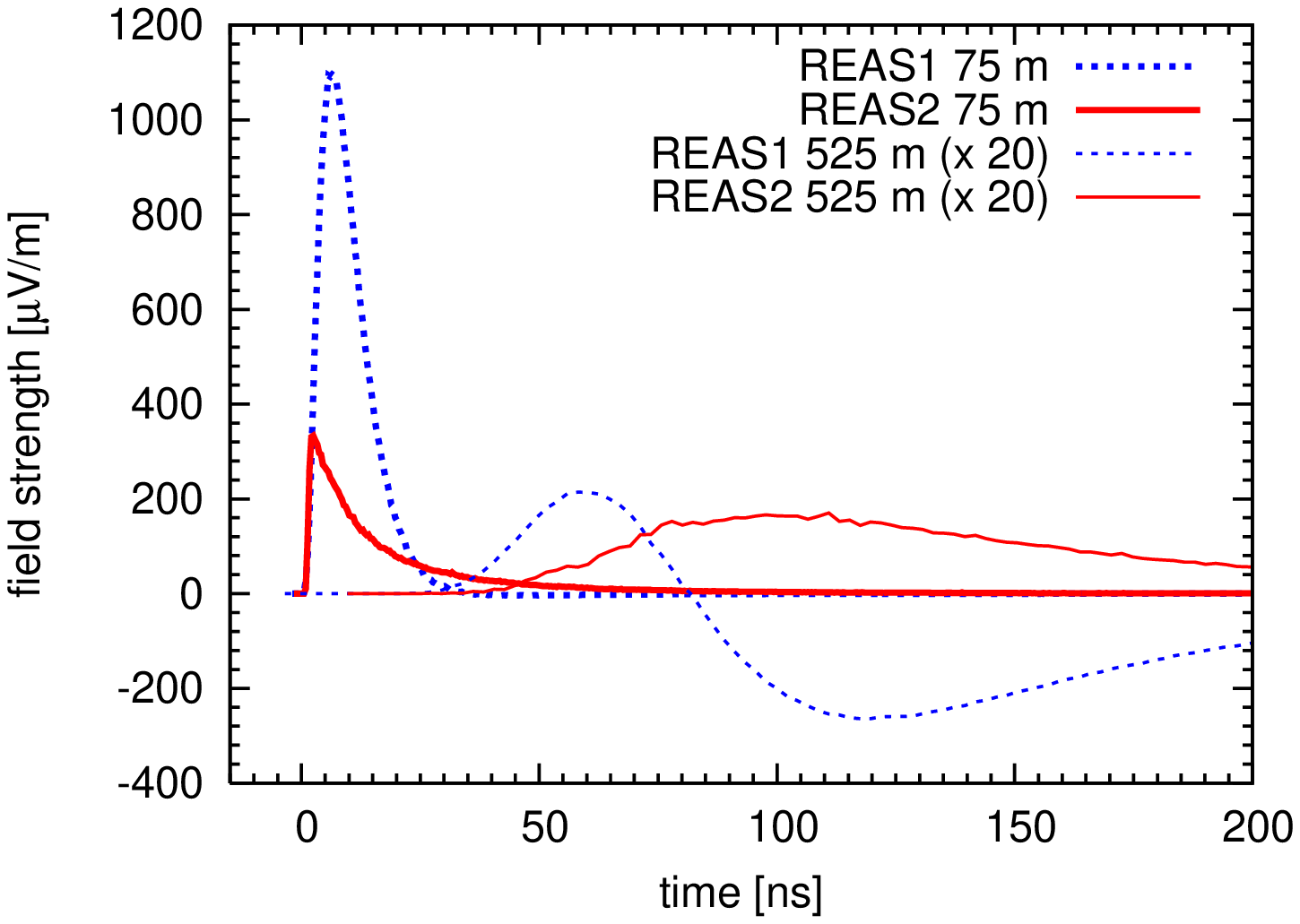}
\caption{\label{overallpulses90deg}Same as Fig.\ \ref{overallpulses} but for observers to the west of the shower centre.\newline}
\end{minipage}
\end{figure}

\begin{figure}[htb]
\begin{minipage}{15.5pc}
\includegraphics[angle=270,width=15.5pc]{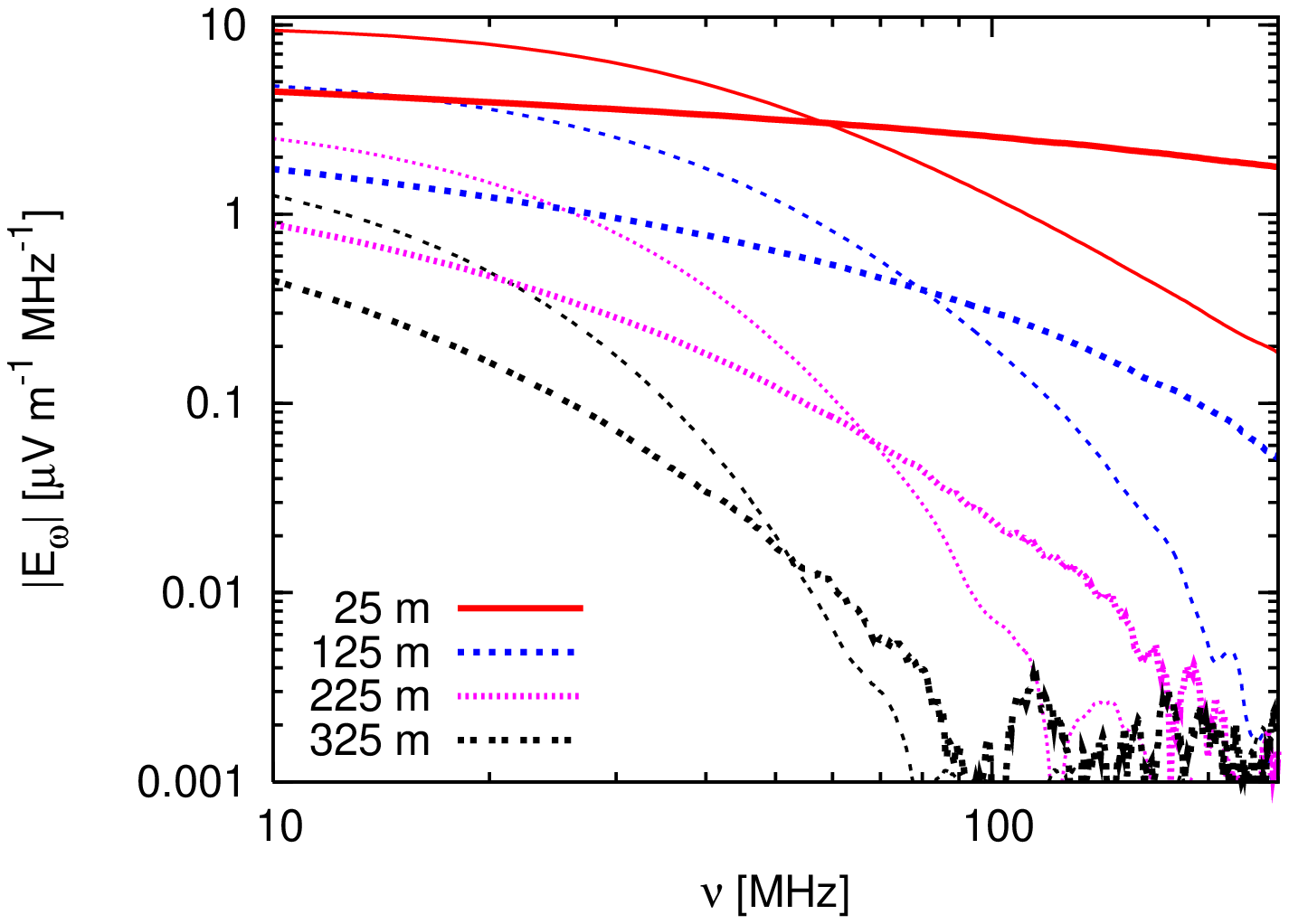}
\caption{\label{overallspectra}Overall change in the radio signal frequency spectra between REAS1 (thin lines) and REAS2 (thick lines) at various distances to the north of the shower centre.}
\end{minipage} \hspace{1.5pc}
\begin{minipage}{15.5pc}
\includegraphics[angle=270,width=15.5pc]{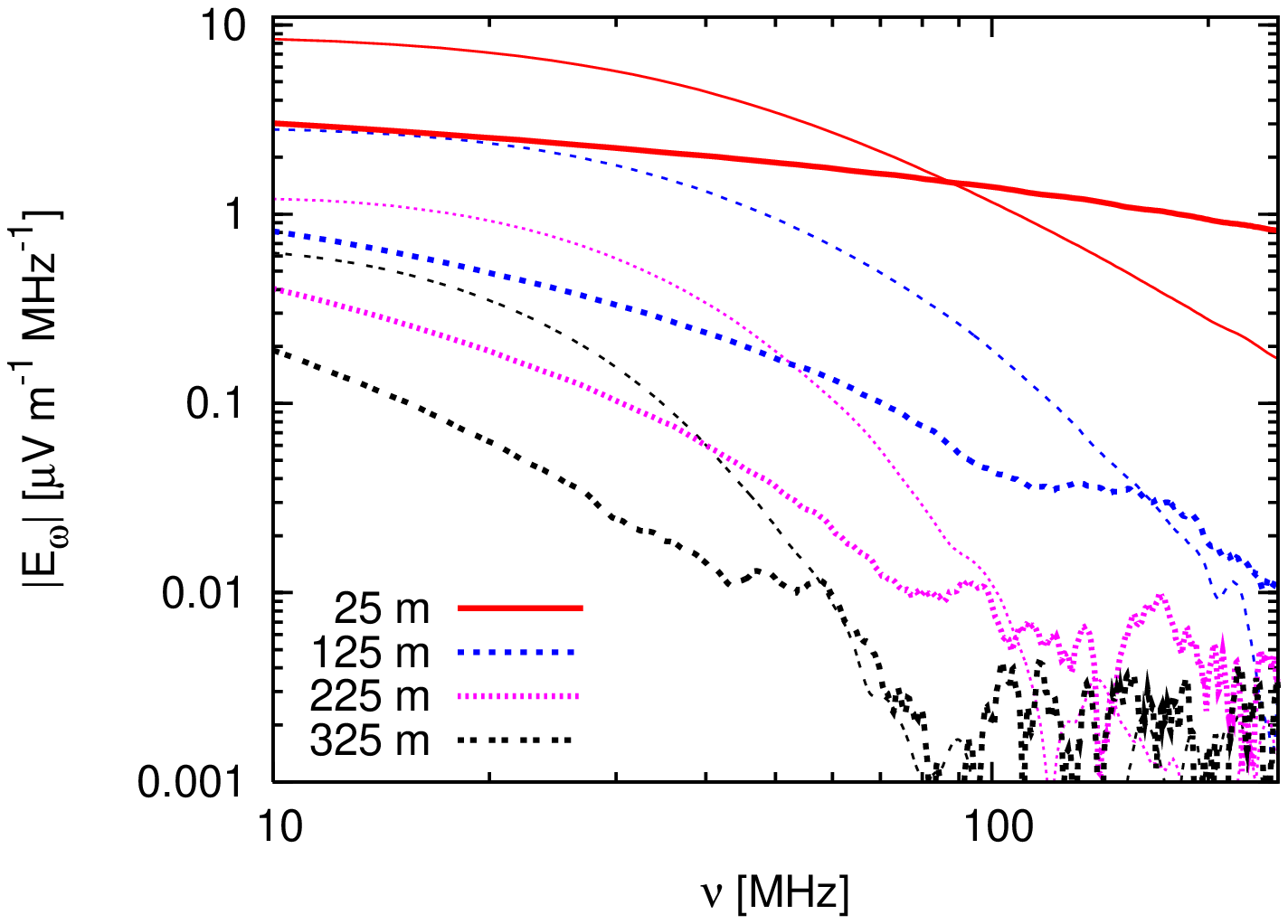}
\caption{\label{overallspectra90deg}Same as Fig.\ \ref{overallspectra} but for observers to the west of the shower centre.\newline\newline}
\end{minipage}
\end{figure}

Figure \ref{overallpulses} shows the overall difference in the calculated radio pulses for observers along the north-south axis from the shower centre. The pulses are damped up to a factor of two in amplitude. Pulses close to the shower centre become much narrower with a much steeper rising edge. This is also reflected in the spectra shown in Fig.\ \ref{overallspectra}: close to the shower axis, the spectra are now much flatter. In the observing bandwidth of the LOPES experiment (40--80~MHz), the frequency-averaged field strength is changed only very little.

The situation is more complex for observers along the east-west axis from the shower centre as shown in Fig.\ \ref{overallpulses90deg}. Observers close to the shower axis receive a strongly damped radio pulse. For observers at larger distances the pulse amplitude changes much less, but the bipolarity of the pulses vanishes as a consequence of the more realistic particle distributions.

\begin{figure}[htb]
\begin{minipage}{15.5pc}
\includegraphics[angle=270,width=15.5pc]{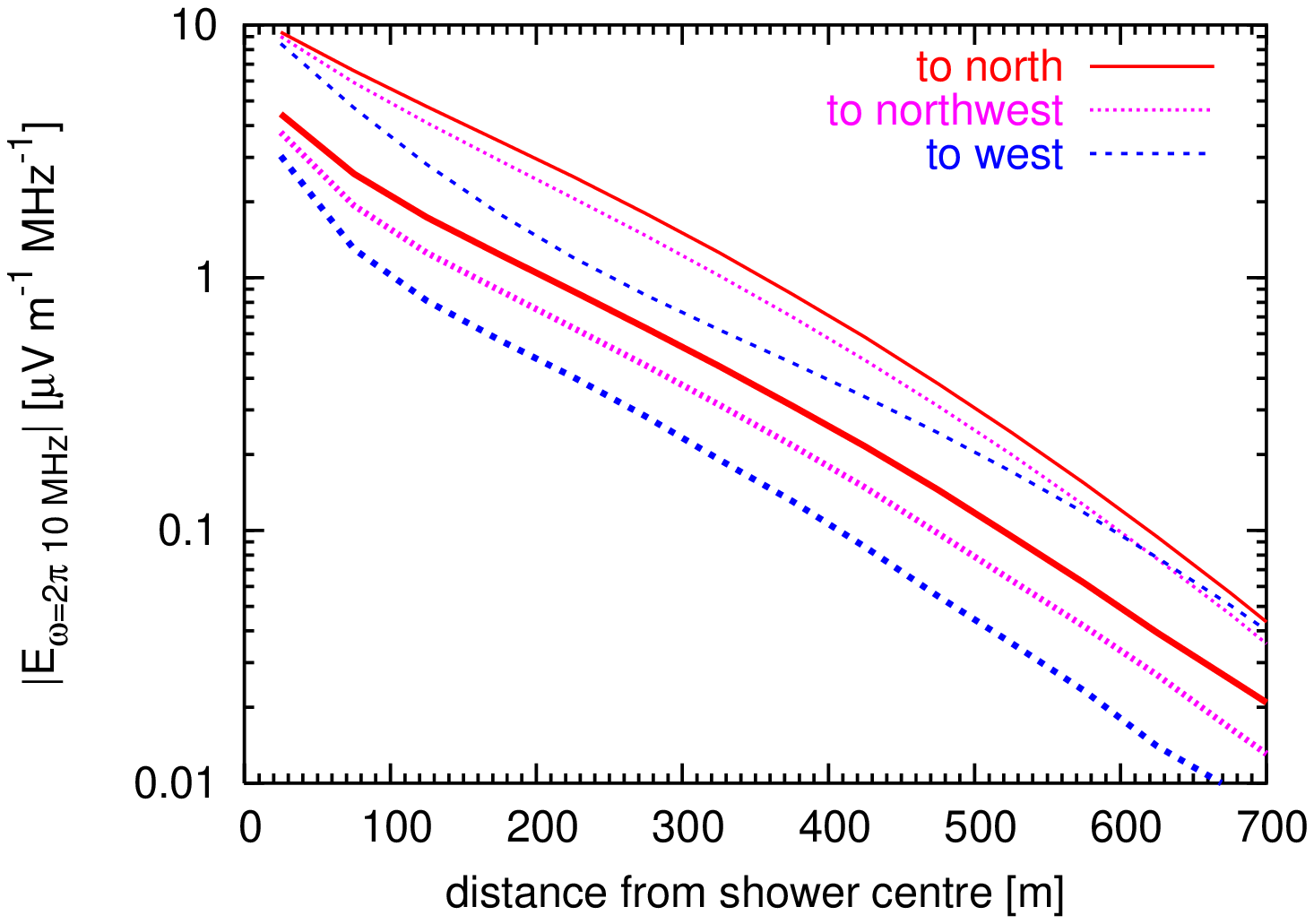}
\caption{\label{radial}Spectral field strength at 10~MHz as a function of radial distance to the shower centre for different azimuthal directions compared between REAS1 (thin lines) and REAS2 (thick lines).}
\end{minipage} \hspace{1.5pc}
\begin{minipage}{15.5pc}
\includegraphics[angle=270,width=15.5pc]{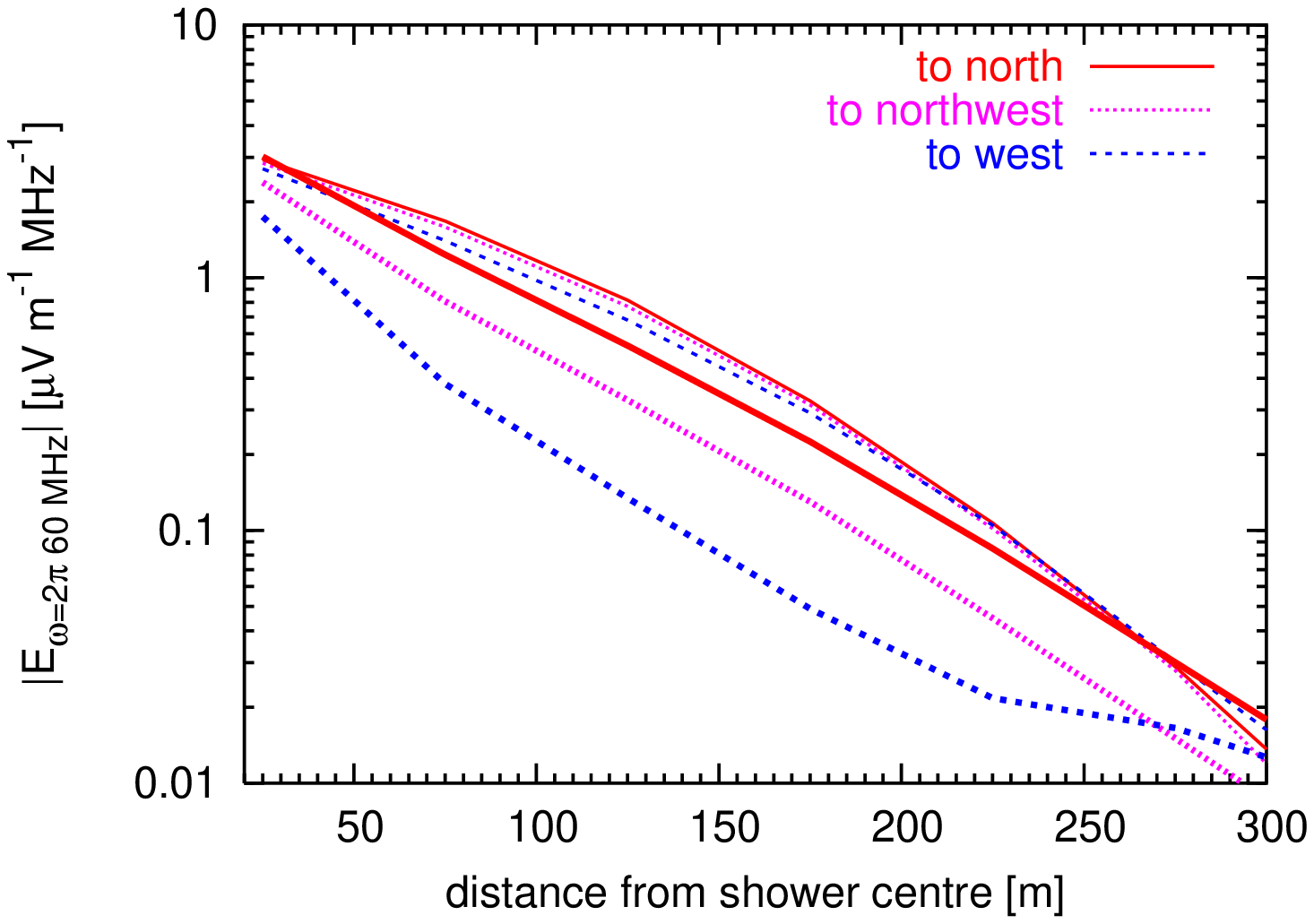}
\caption{\label{radial60MHz}Same as Fig.\ \ref{radial} for the 60~MHz frequency component.\newline\newline\newline\newline}
\end{minipage}
\end{figure}

Figure \ref{radial} illustrates the changes in the lateral distribution of the 10~MHz spectral radio emission component. As expected, there is an overall damping. Except in the centre region, the field strength still decreases approximately exponentially with distance, and the scale parameter of $\approx 125$~m for the emission from the histogrammed shower is not much different from that derived for the parametrised shower. However, a more pronounced asymmetry between north-south and east-west direction is observed. At 60~MHz (shown in Fig.\ \ref{radial60MHz}) the overall drop in amplitude is less pronounced than at 10 MHz due to the now flatter frequency spectra. The scale parameter there corresponds to $\approx 50$~m, but the exponential decrease is only valid up to $\sim250$--300~m distance.

\begin{figure}[!ht]
   \centering
   \includegraphics[width=4.1cm,angle=270]{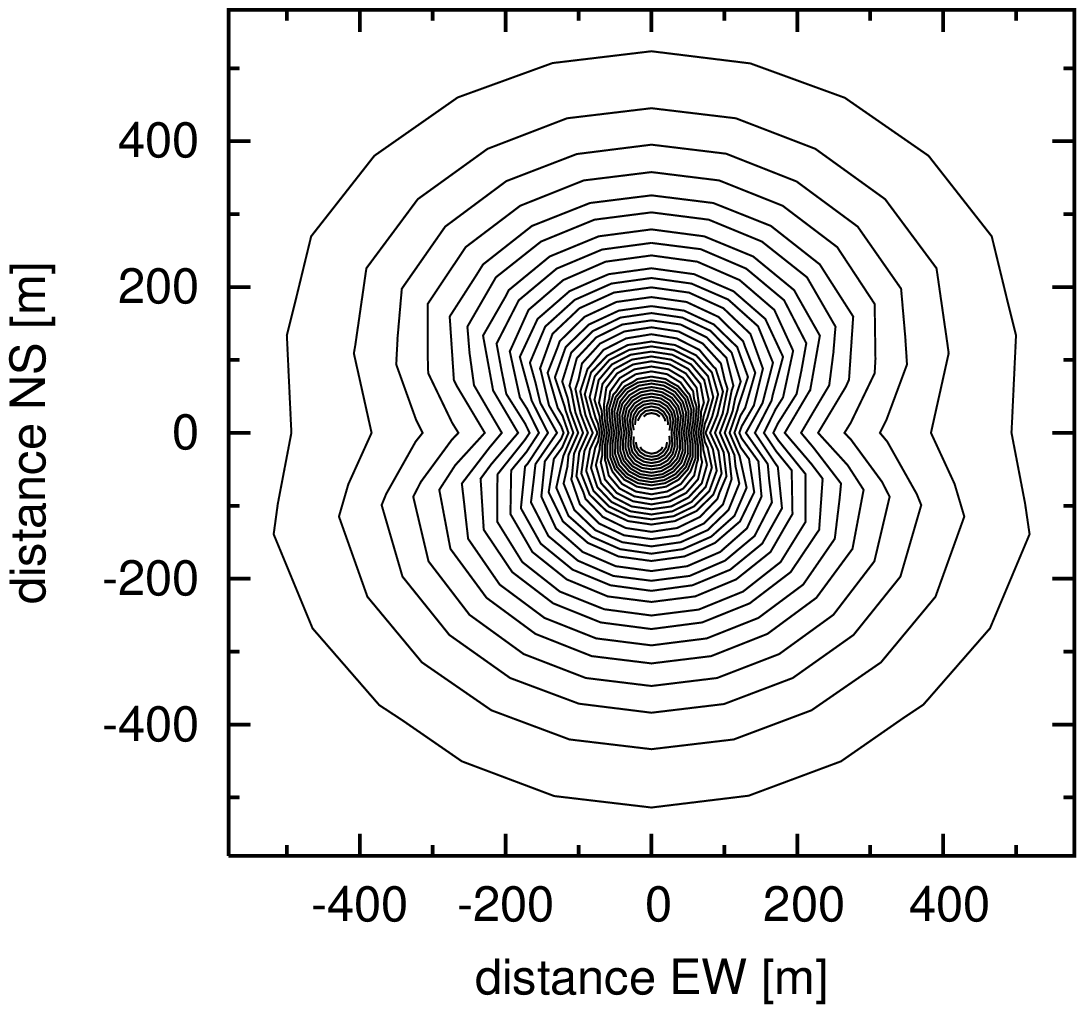}
   \includegraphics[width=4.1cm,angle=270]{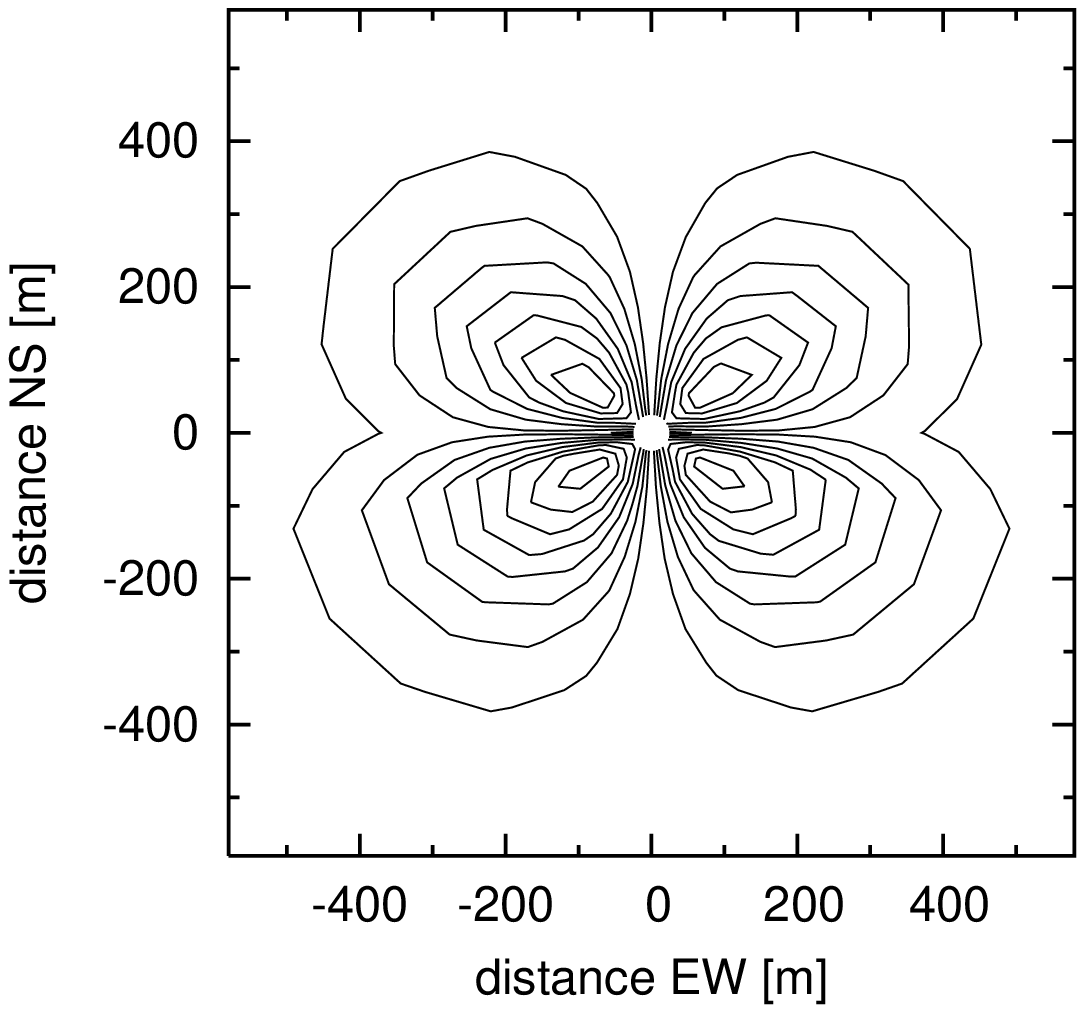}
   \includegraphics[width=4.1cm,angle=270]{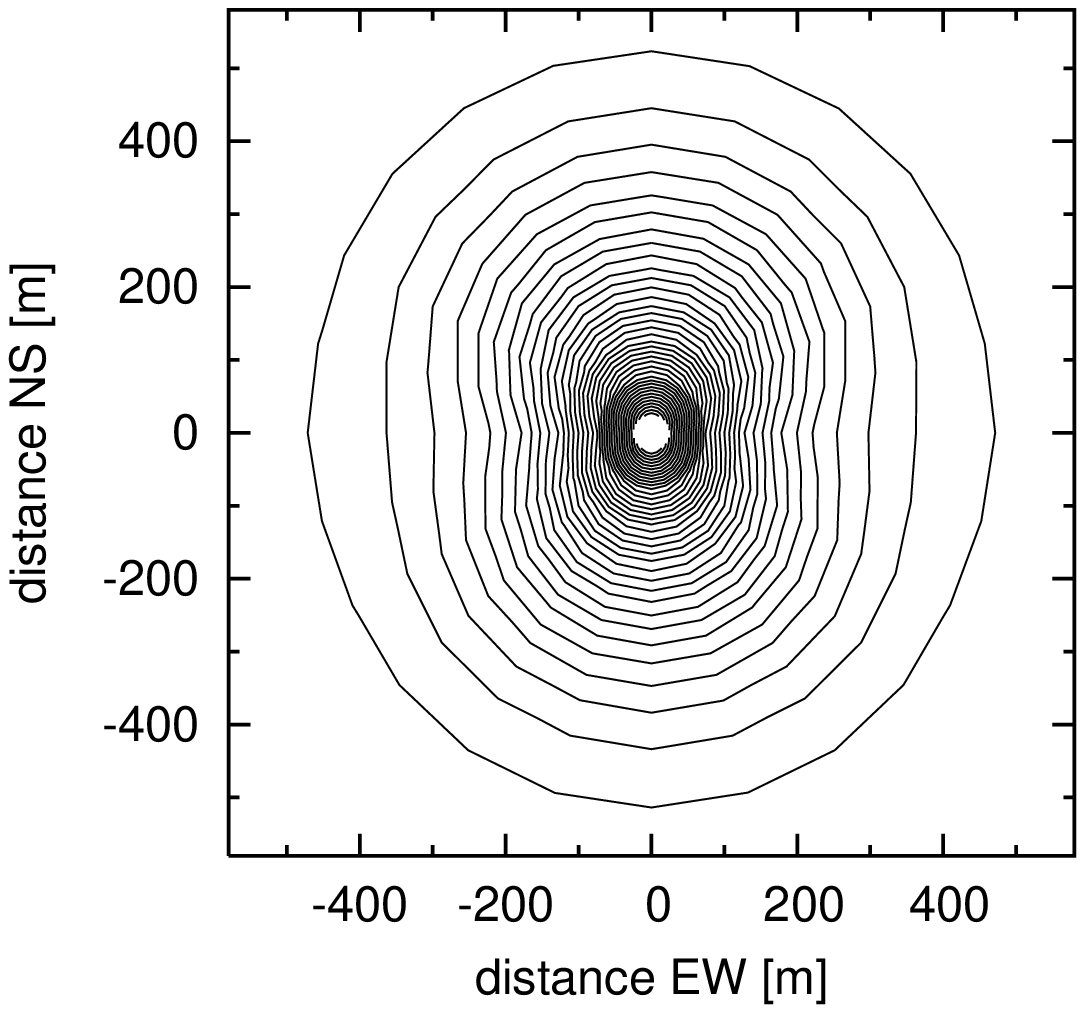}\\
   \includegraphics[width=4.1cm,angle=270]{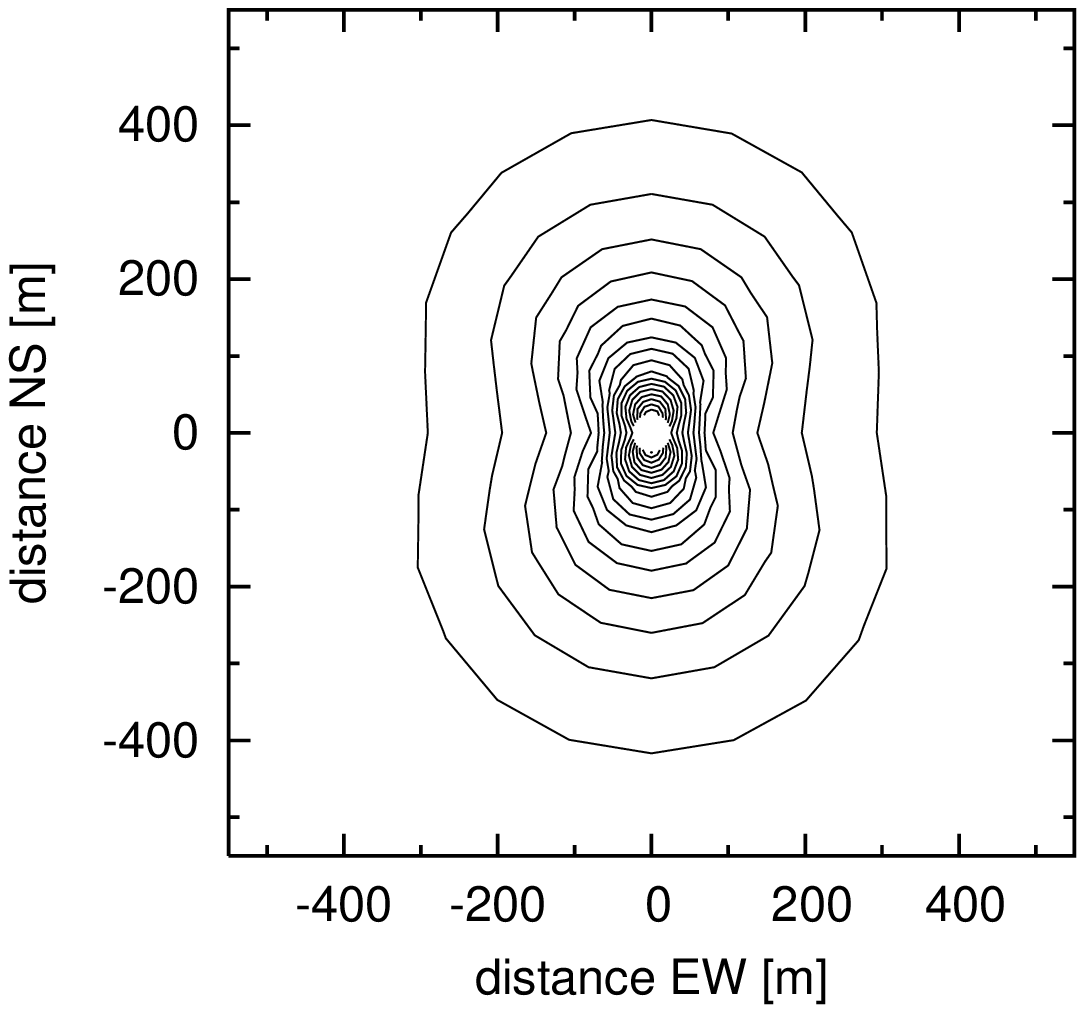}
   \includegraphics[width=4.1cm,angle=270]{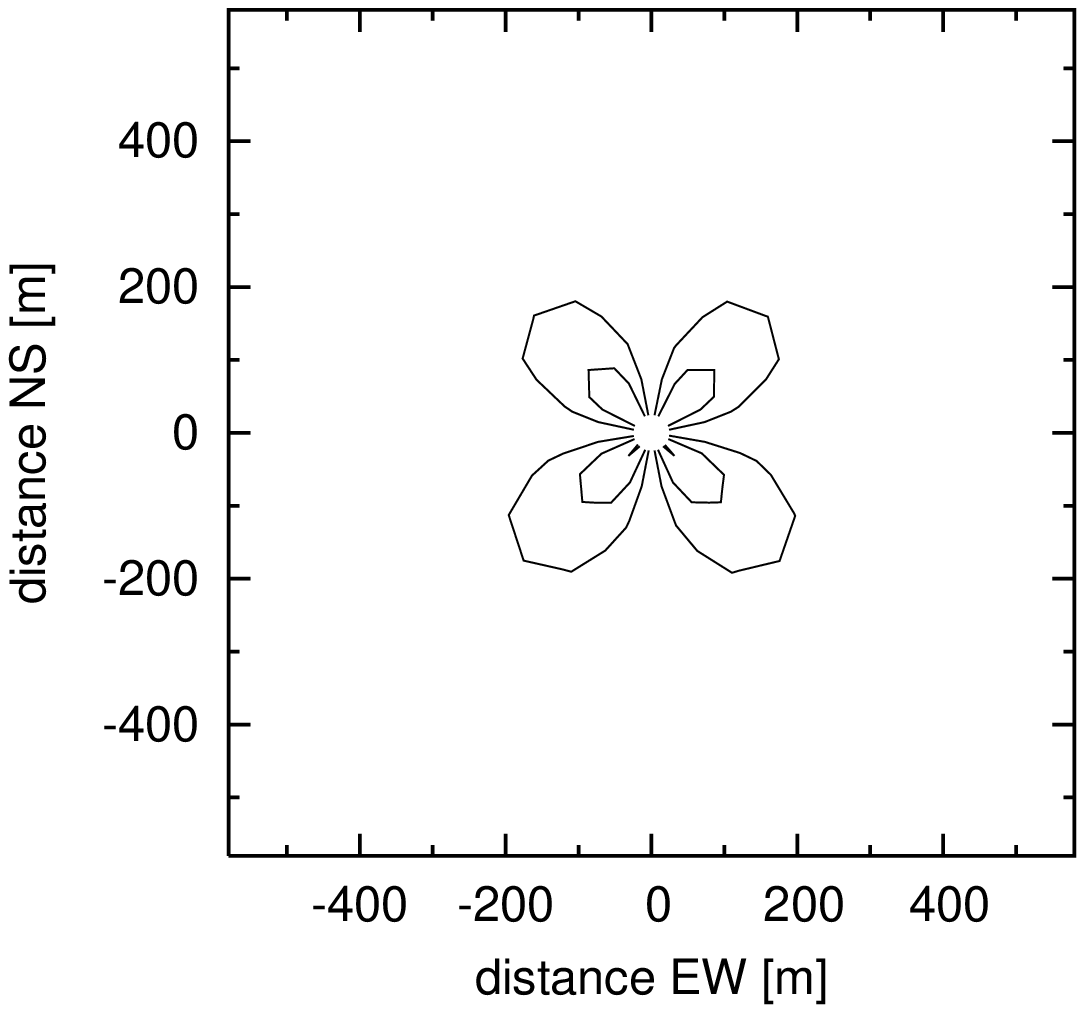}
   \includegraphics[width=4.1cm,angle=270]{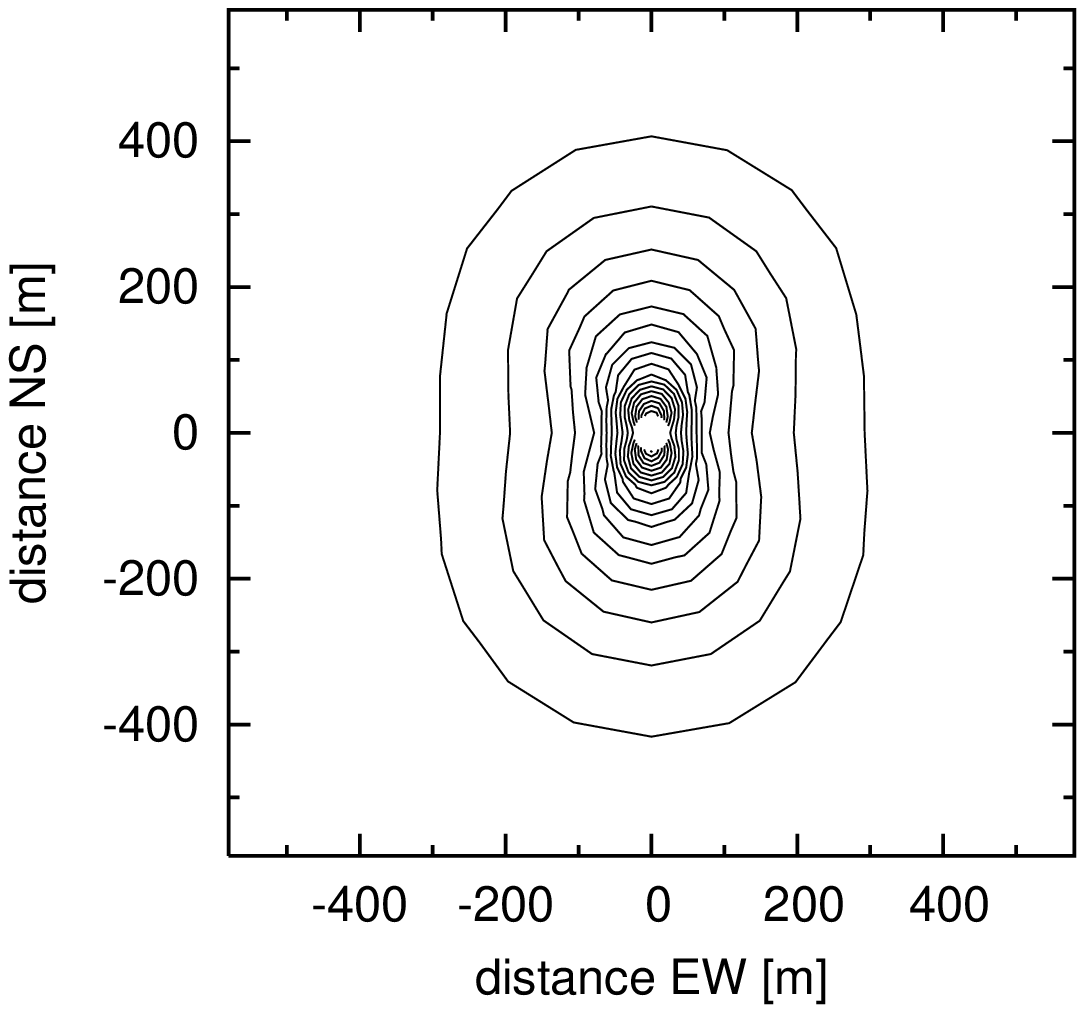}\\
   \caption[Contour plots of the parametrised and histogrammed air shower.]{
   \label{contoursvorhernachher}
   Contour plots of the parametrised (upper) and histogrammed (lower) air shower emission at $\nu=10$~MHz. The columns (from left to right) show the total field strength, the north-south polarisation component and the east-west polarisation component. The vertical polarisation component (not shown here) does not contain any significant flux. Contour levels are 0.25~$\mu$V~m$^{-1}$~MHz$^{-1}$ apart in $E_{\omega}$, outermost contour corresponds to 0.25~$\mu$V~m$^{-1}$~MHz$^{-1}$. White centre region has not been calculated.
   }
   \end{figure}

The spatial distribution of the 10~MHz radiation component and its polarisation properties are also illustrated by Fig.\ \ref{contoursvorhernachher}. One can easily identify the overall damping of the emission. In addition, the radio ``footprint'' becomes more asymmetric and less circular. The polarisation characteristics remain qualitatively very similar, even though the realistic electron to positron ratio is now taken into account.

%______________________________________________________________
% 4.1 cm for normal style, 3.0 cm for referee style
%______________________________________________________________

\begin{figure}[!ht]
   \centering
   \includegraphics[width=4.1cm,angle=270]{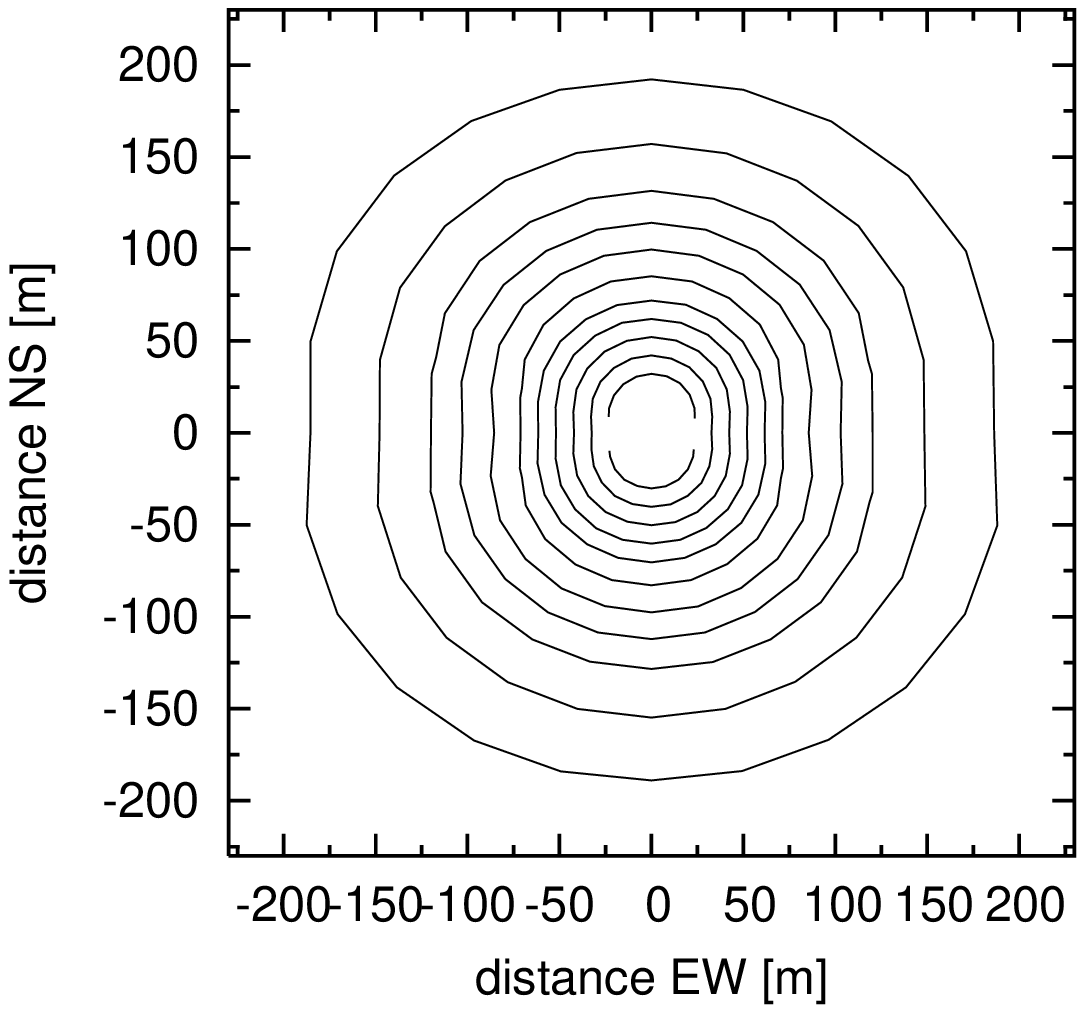}
   \includegraphics[width=4.1cm,angle=270]{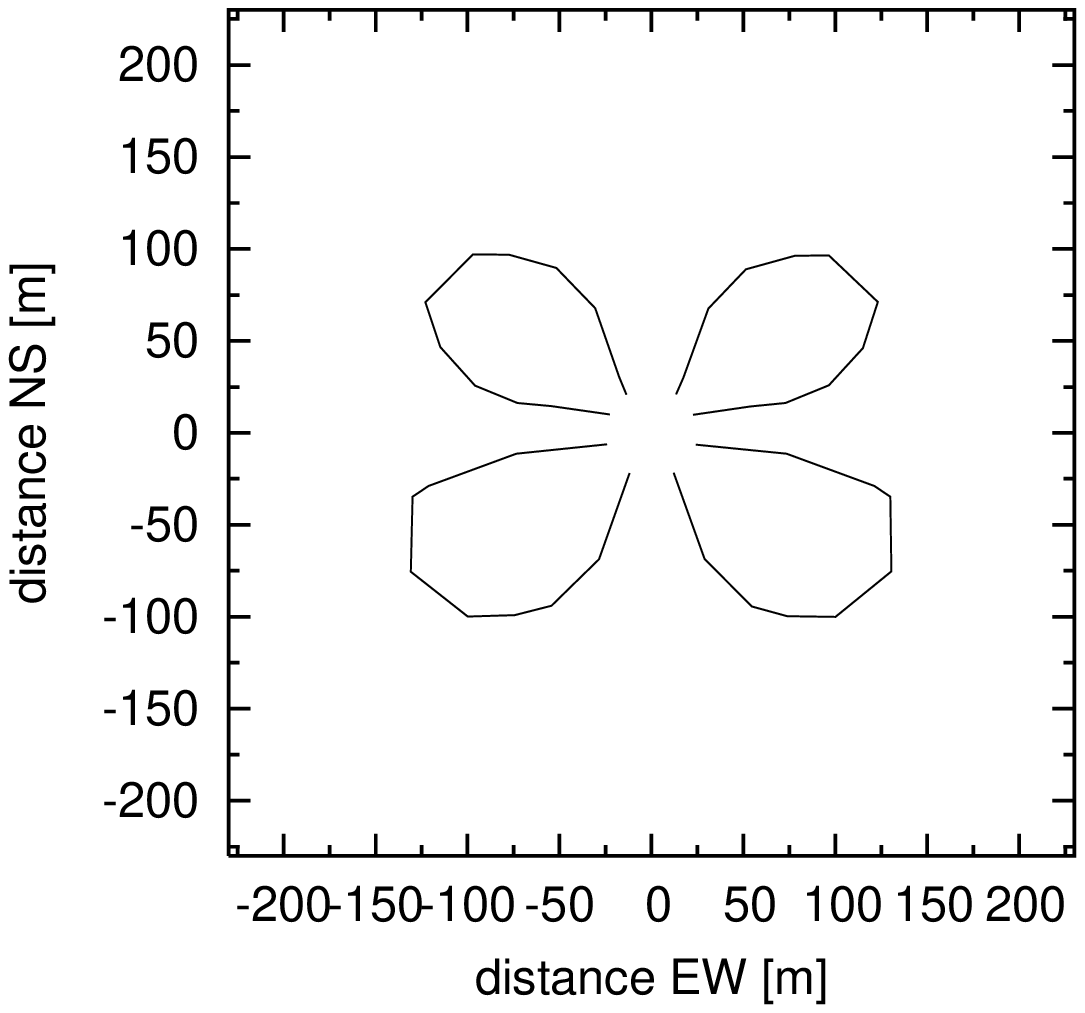}
   \includegraphics[width=4.1cm,angle=270]{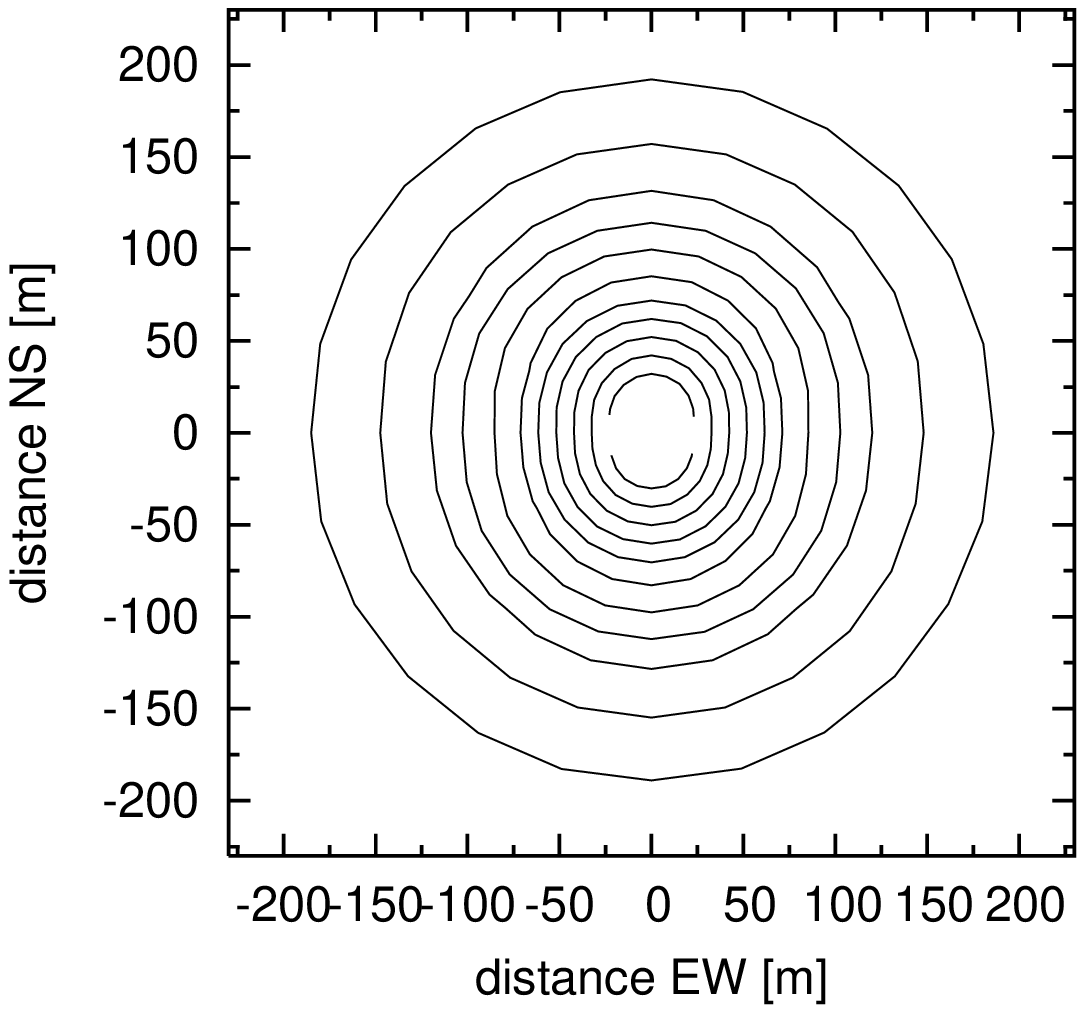}\\
   \includegraphics[width=4.1cm,angle=270]{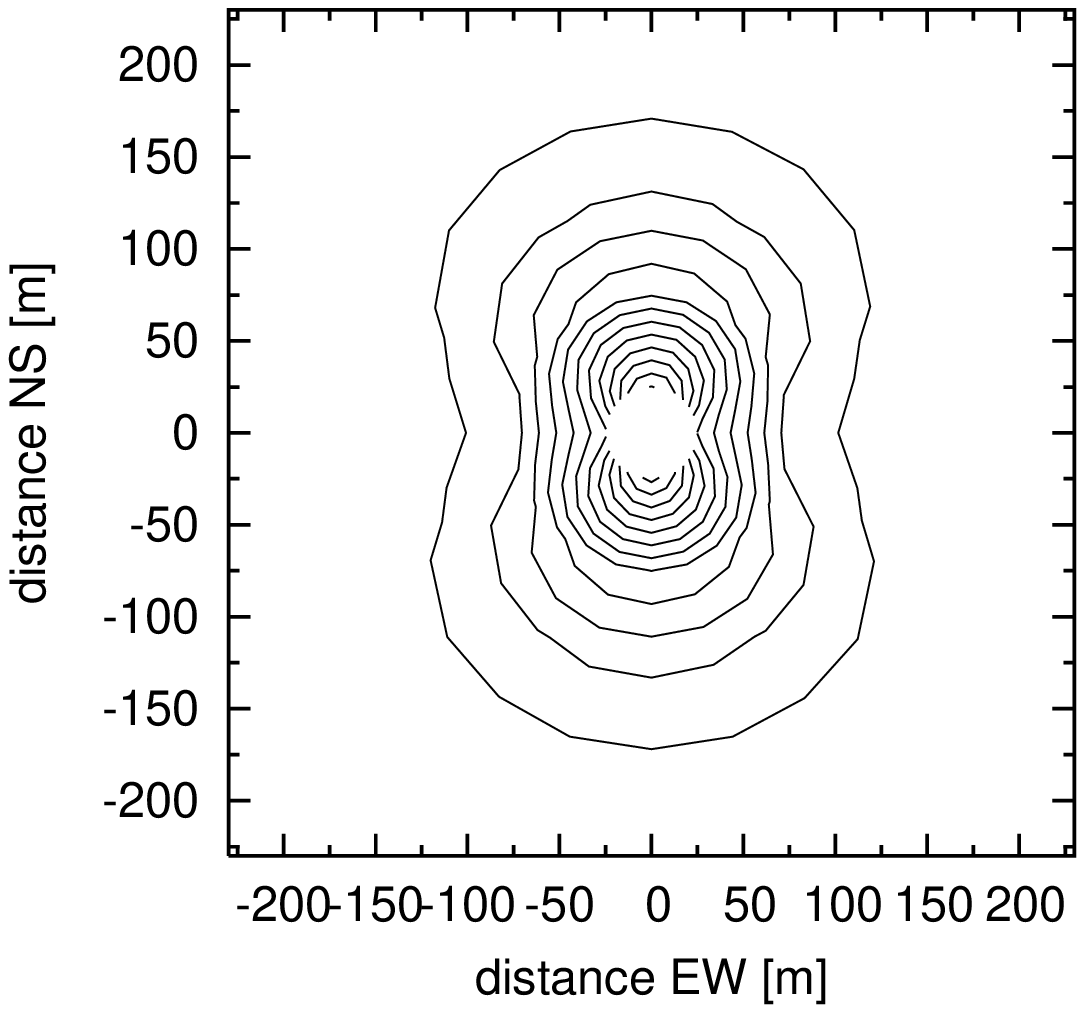}
   \includegraphics[width=4.1cm,angle=270]{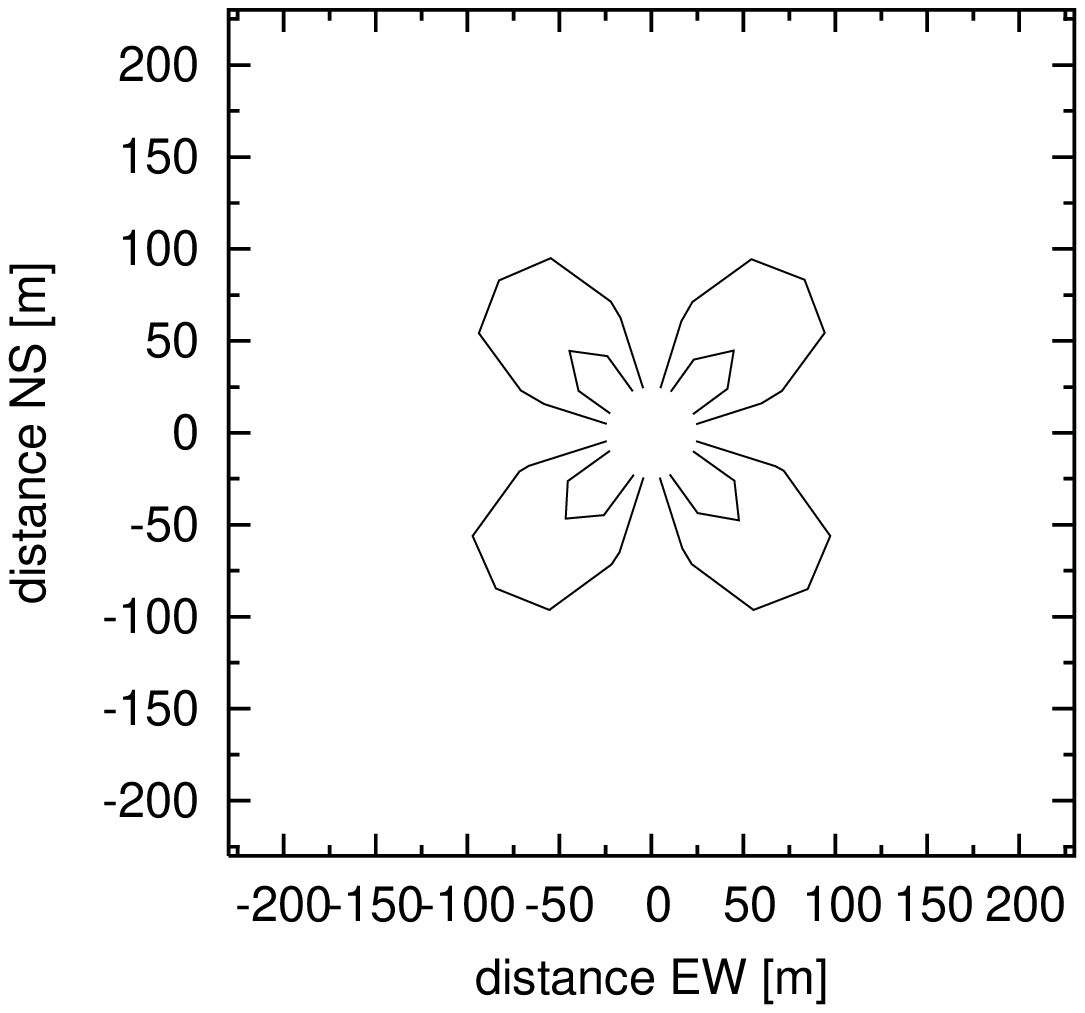}
   \includegraphics[width=4.1cm,angle=270]{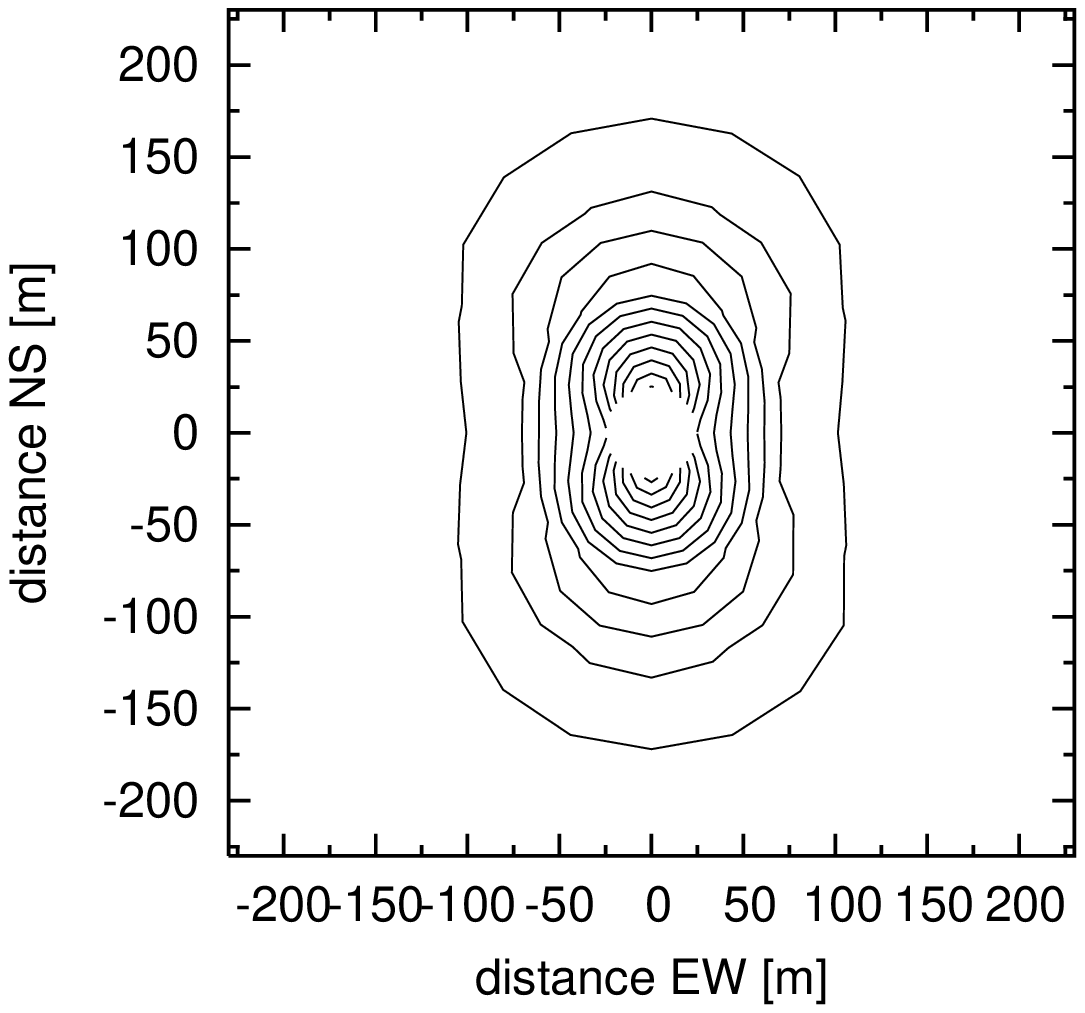}\\
   \caption[Contour plots of the parametrised and histogrammed air shower.]{
   \label{contoursvorhernachher60MHz}
   Same as Fig.\ \ref{contoursvorhernachher} but for $\nu=60$~MHz observing frequency.
   }
   \end{figure}

Figure \ref{contoursvorhernachher60MHz} shows the changes occurring at 60~MHz, the central frequency of the LOPES observing bandwidth. As a consequence of the flatter spectra, the changes at 60~MHz are much smaller than at 10~MHz. The increasing asymmetry of the footprint, however, remains.

%______________________________________________________________
% 4.1 cm for normal style, 3.0 cm for referee style
%______________________________________________________________

\section{Pulse shape analysis} \label{pulseshapeanalysis}

In REAS2, detailed information on the relevant particle distributions and their development throughout the full air shower evolution is available. This information can be used to analyse how different regimes identifiable in the particle distributions contribute to the radio pulses. In the following we discuss a few examples of such analyses.

\subsection{Particle energy}

\begin{figure}[htb]
\begin{minipage}{15.5pc}
\includegraphics[angle=270,width=15.5pc]{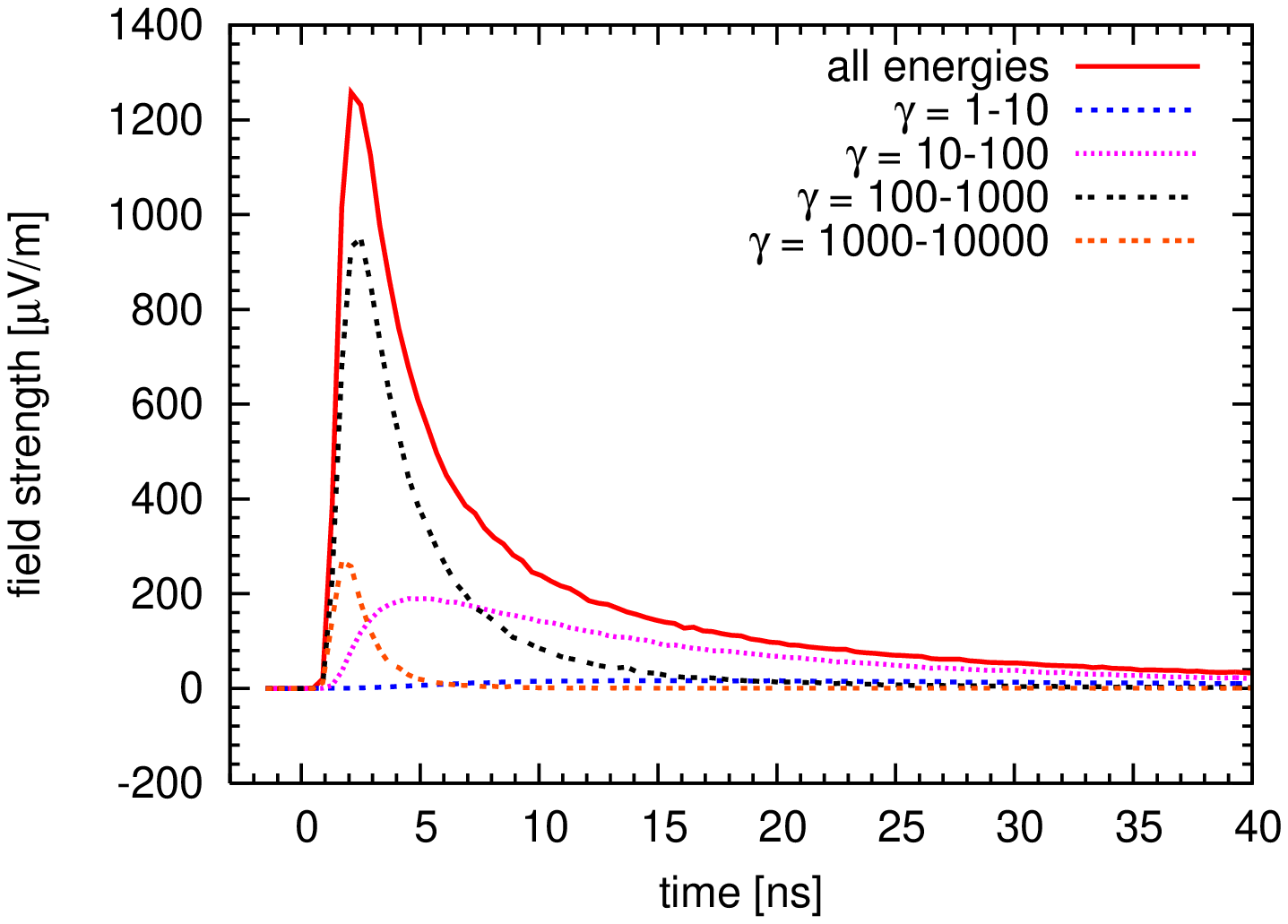}
\caption{\label{energyregimes75m}Contribution of different particle energy regimes to the radio pulse at 75~m north from the shower centre.}
\end{minipage} \hspace{1.5pc}
\begin{minipage}{15.5pc}
\includegraphics[angle=270,width=15.5pc]{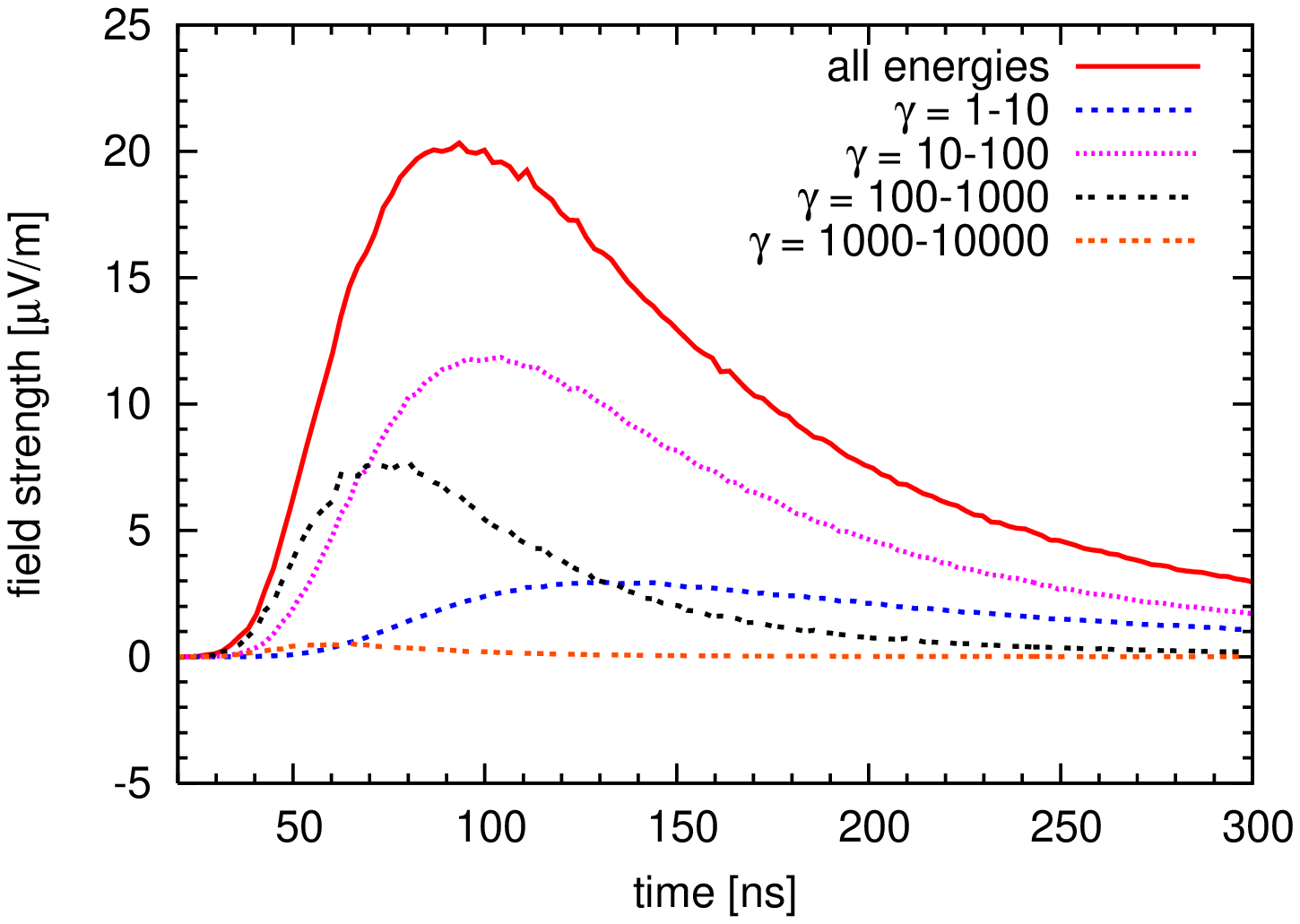}
\caption{\label{energyregimes525m}Contribution of different particle energy regimes to the radio pulse at 525~m north from the shower centre.}
\end{minipage}
\end{figure}

Figure \ref{energyregimes75m} shows the contribution of different particle energy regimes to the overall radio pulse close to the shower centre. It is evident that high-energy particles with Lorentz factors between 100 and 1000 dominate the emission, although low-energy particles outnumber them by far (cf.\ Fig.\ \ref{energyhisto}). The reason for this is that high-energy particles radiate more efficiently than low-energy particles, but only into narrow beaming-cones and only from close to the shower axis where they are prominent. Consequently, the pulse at 525~m is dominated by the lower energy particles with Lorentz factors between 10 and 100, as shown in Fig.\ \ref{energyregimes525m}.

\subsection{Lateral distance} \label{rings}

\begin{figure}[htb]
\begin{minipage}{15.5pc}
\includegraphics[angle=270,width=15.5pc]{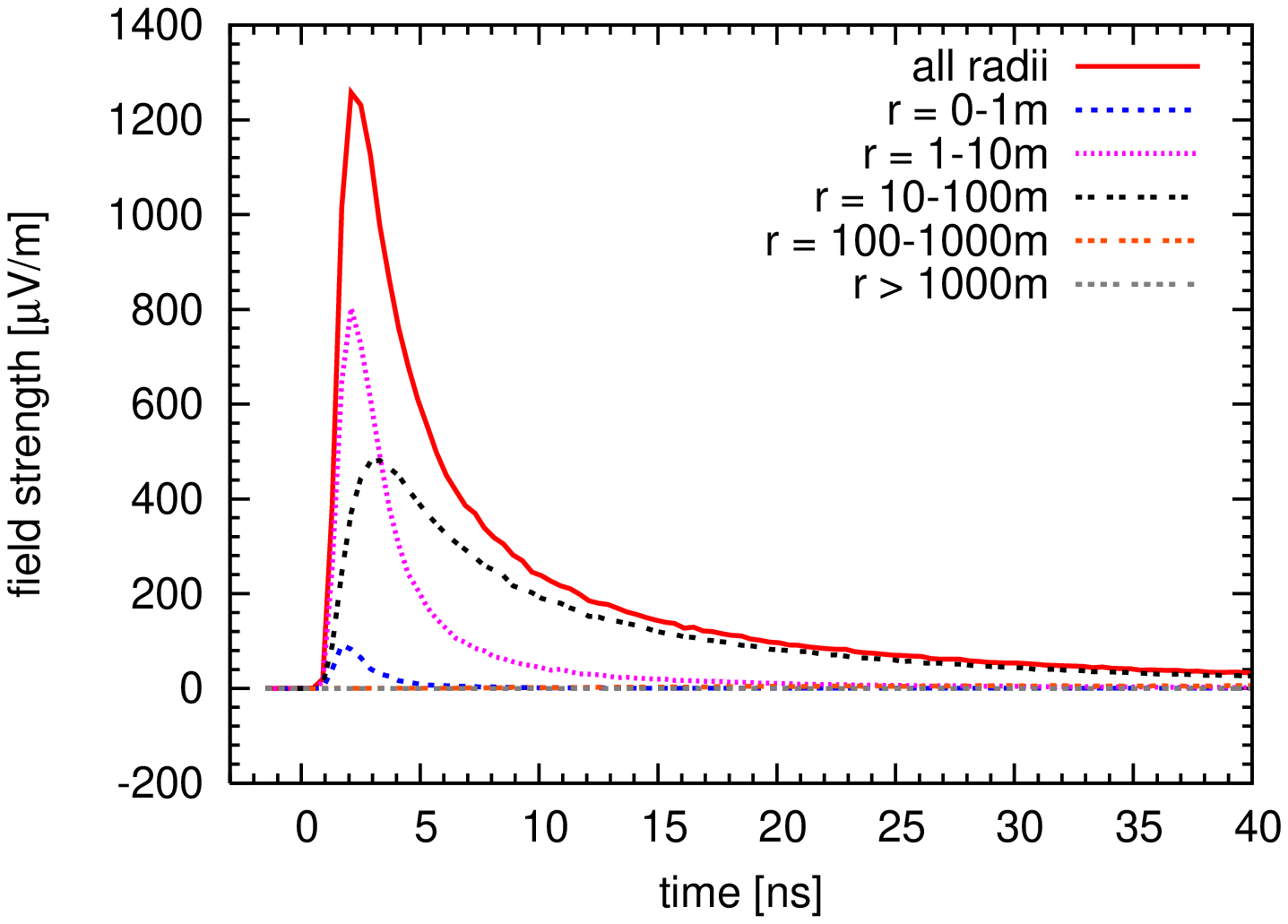}
\caption{\label{radiusregimes75m}Contribution of different lateral rings to the radio pulse at 75~m north from the shower centre.}
\end{minipage} \hspace{1.5pc}
\begin{minipage}{15.5pc}
\includegraphics[angle=270,width=15.5pc]{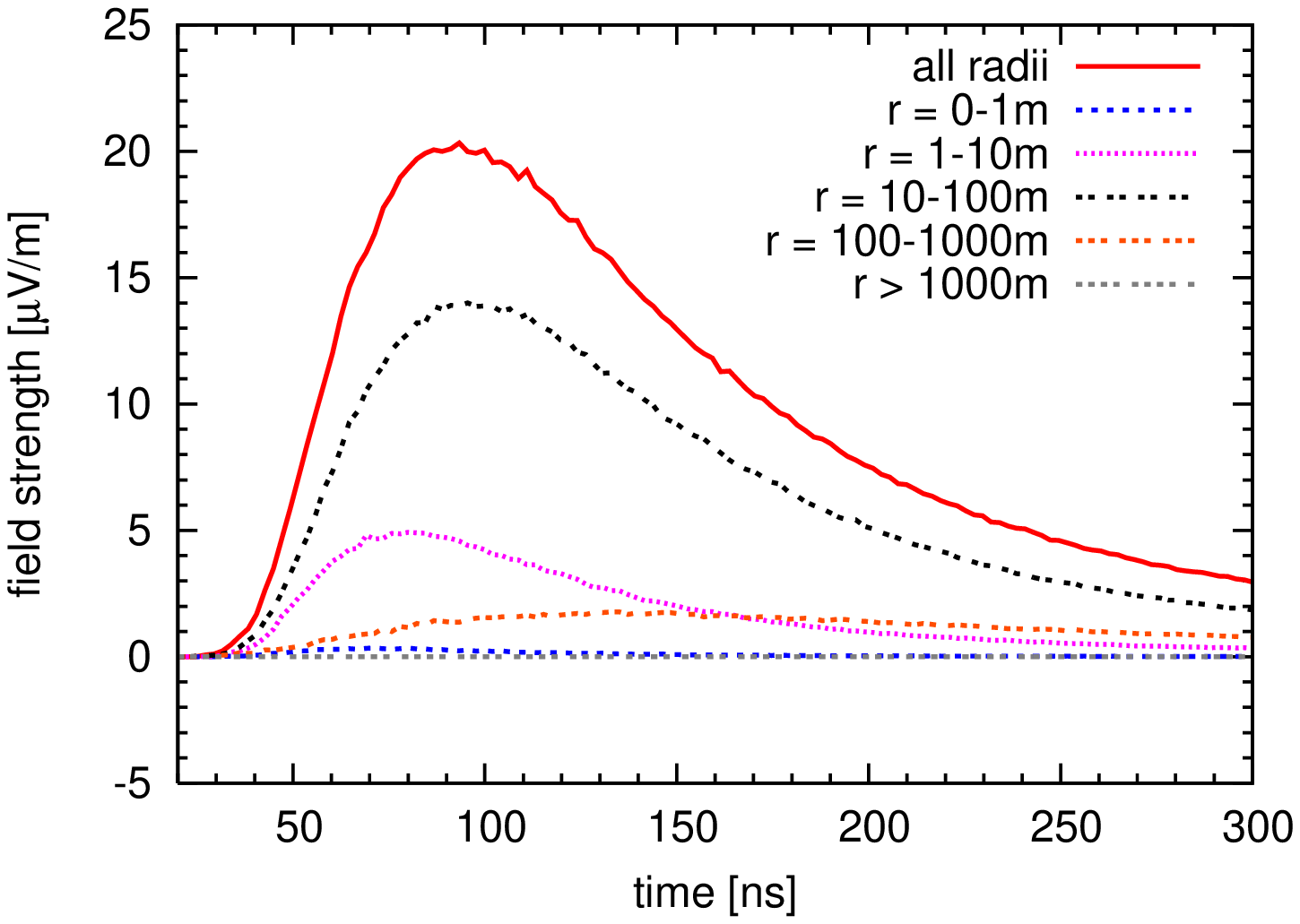}
\caption{\label{radiusregimes525m}Contribution of different lateral rings to the radio pulse at 525~m north from the shower centre.}
\end{minipage}
\end{figure}

Another interesting question is how different lateral distance ranges (i.e., particles in geometrically defined rings, independent of atmospheric depth) contribute to the emission. This is shown in Figures \ref{radiusregimes75m} and \ref{radiusregimes525m}. Close to the shower centre, particles near the shower axis dominate the emission. This is consistent with the result that the high-energy particles (which are clustered close to the shower axis) dominate the emission close to the shower centre. Consequently, at larger distances, the relative importance of the particles close to the shower axis is lower, and lower energy particles with distances of 10--100~m from the axis give the dominant contribution. In all cases, particles at distances of more than 100~m from the axis only contribute very little to the overall pulses.

\subsection{Atmospheric depth}

\begin{figure}[htb]
\begin{minipage}{15.5pc}
\includegraphics[angle=270,width=15.5pc]{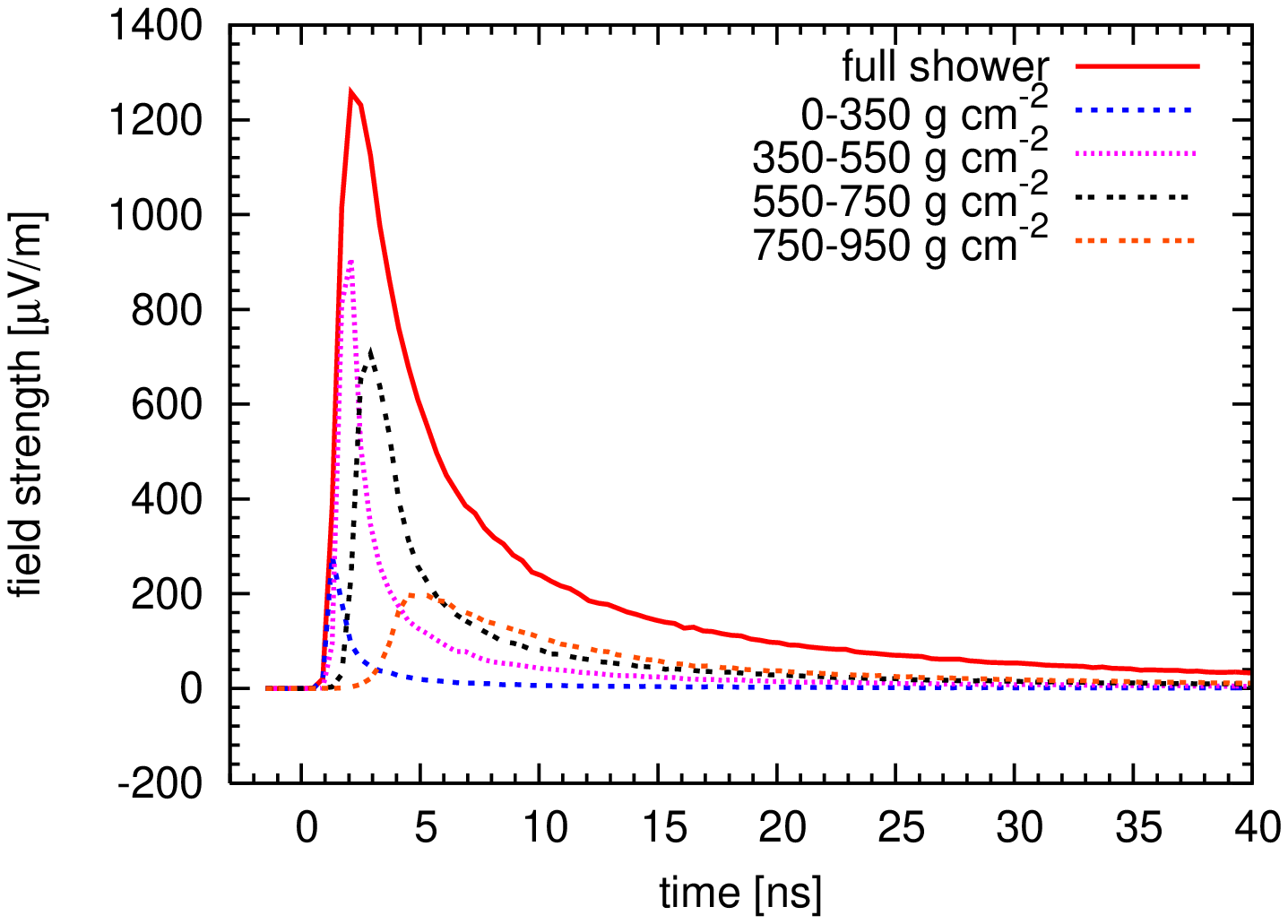}
\caption{\label{depthregimes75m}Contribution of different shower evolution stages to the radio pulse at 75~m north from the shower centre.}
\end{minipage} \hspace{1.5pc}
\begin{minipage}{15.5pc}
\includegraphics[angle=270,width=15.5pc]{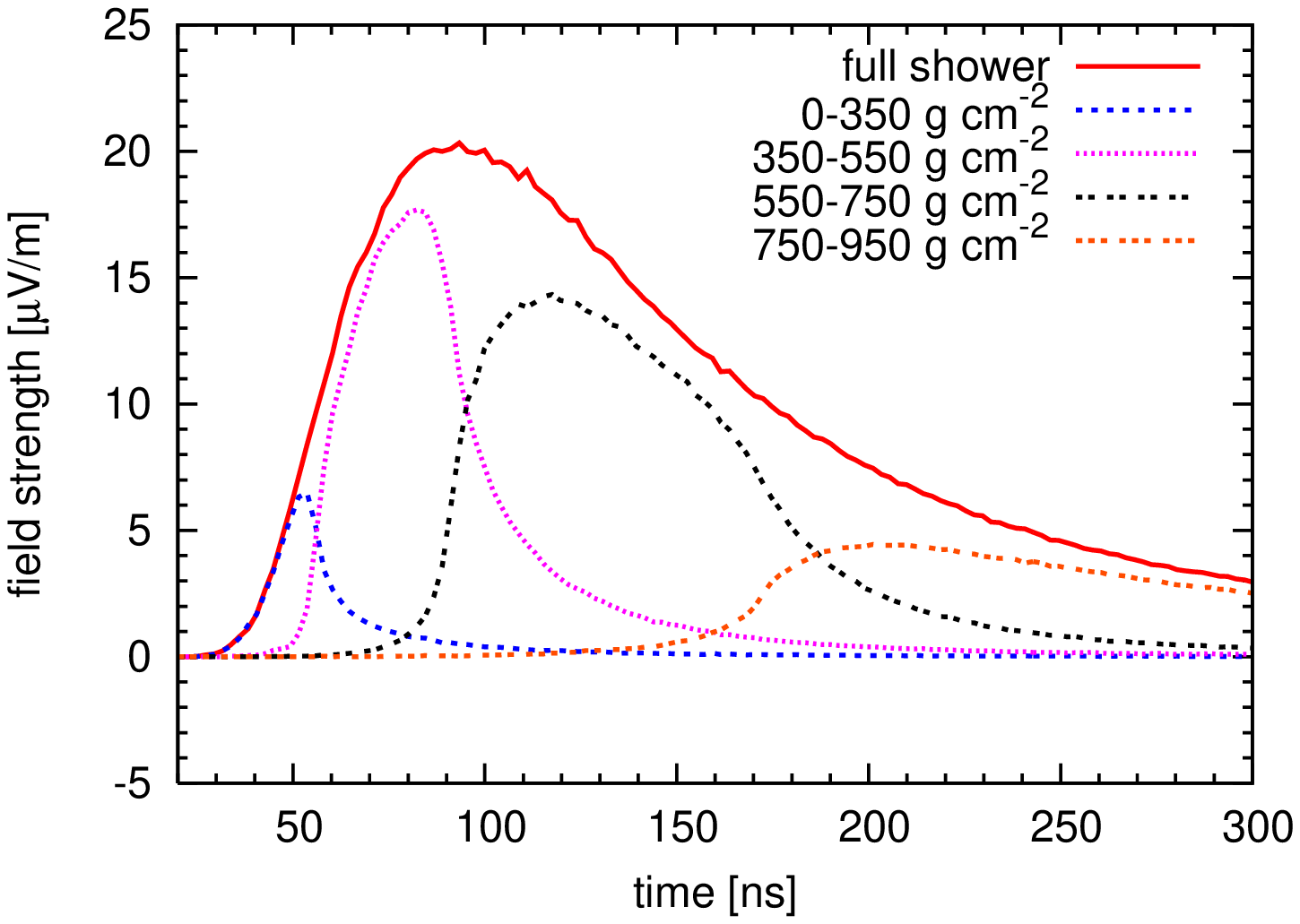}
\caption{\label{depthregimes525m}Contribution of different shower evolution stages to the radio pulse at 525~m north from the shower centre.}
\end{minipage}
\end{figure}

A particularly interesting question is how different stages of the air shower evolution contribute to the radio pulses. Figure \ref{depthregimes75m} illustrates that the radio emission from different slices in atmospheric depth arrive nearly simultaneously at an observer close to the shower centre. The reason is that geometrical time-delays are small at these radial distances and that both the radio emission and the leading particle travel with the speed of light in the current simulations. If the refractive index profile of the atmosphere is taken into account, the contributions are likely to arrive at slightly different times and the overall pulse will get slightly broader.

At larger distances from the shower axis, the relative arrival times between the emission from different air shower stages are mainly governed by geometrical time-delays. Figure \ref{depthregimes525m} shows the radio pulse at 525~m to the north from the shower centre. One can clearly differentiate the contributions from the different air shower stages. The emission is dominated by the shower stage around the shower maximum (which here is at $\sim640$~g~cm$^{-2}$) and the stage shortly before the shower maximum, whereas the stage after the shower maximum contributes less radiation. This is somewhat surprising as the particle numbers before and after the shower maximum are similar but the particles after the shower maximum are closer to the observer and thus could have been expected to contribute more strongly. Two effects, however, overcompensate this simple distance dependence: For once, the emission is strongly beamed. Emission from earlier shower stages is thus distributed over a larger area and therefore leads to stronger radio signals for observers not very close to the shower centre. In addition, the particles in the stage before the shower maximum traverse a much thinner medium than those in later stages and thus propagate for longer (geometrical) distances, during which they can emit radio signals. In general, a thinner medium leads to higher geosynchrotron radio emission per traversed depth in g~cm$^{-2}$ --- if the shower is able to evolve to high particle multiplicities.

\subsection{Geometric height}

\begin{figure}[htb]
\begin{minipage}{15.5pc}
\includegraphics[angle=270,width=15.5pc]{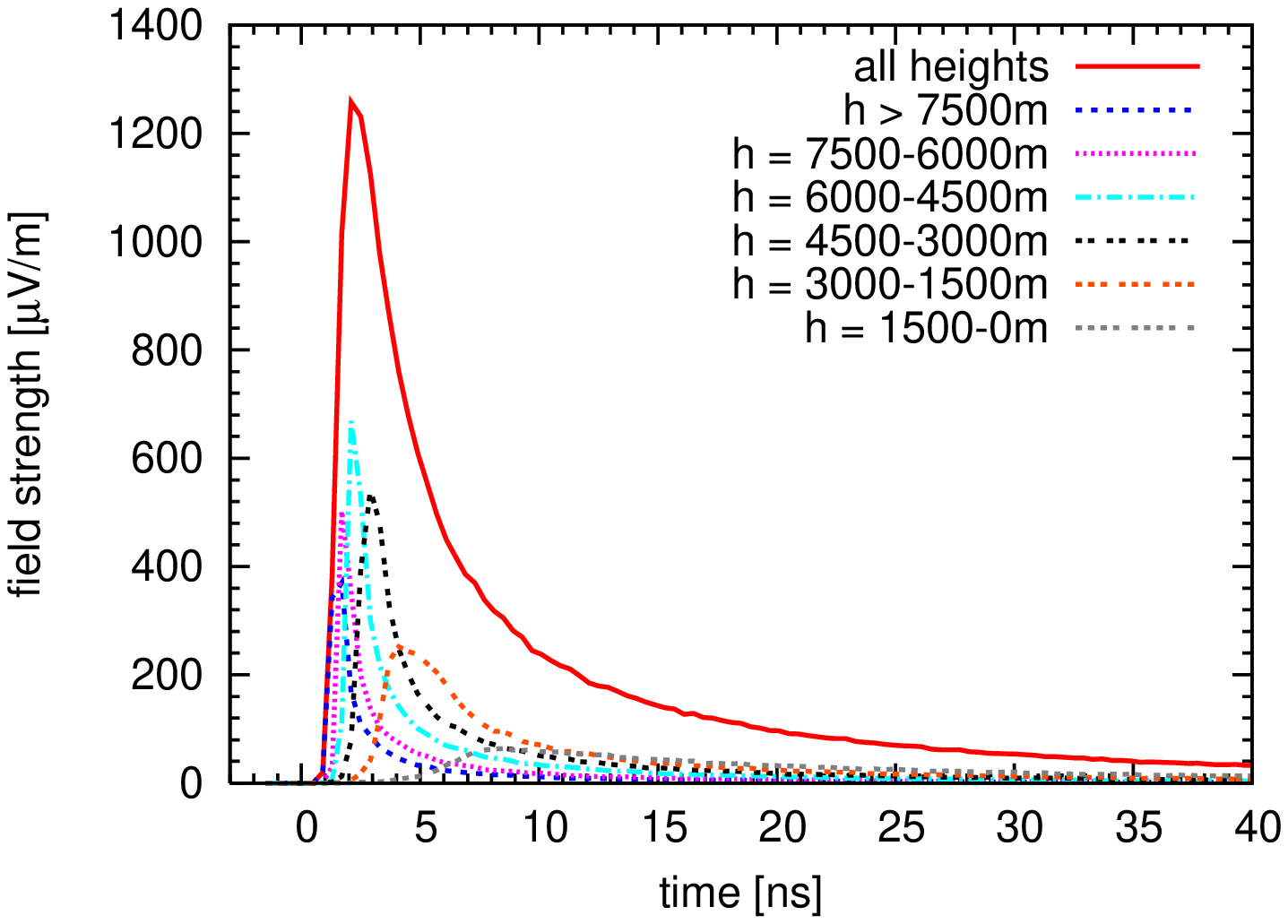}
\caption{\label{heightregimes75m}Contribution of different heights to the radio pulse at 75~m north from the shower centre.}
\end{minipage} \hspace{1.5pc}
\begin{minipage}{15.5pc}
\includegraphics[angle=270,width=15.5pc]{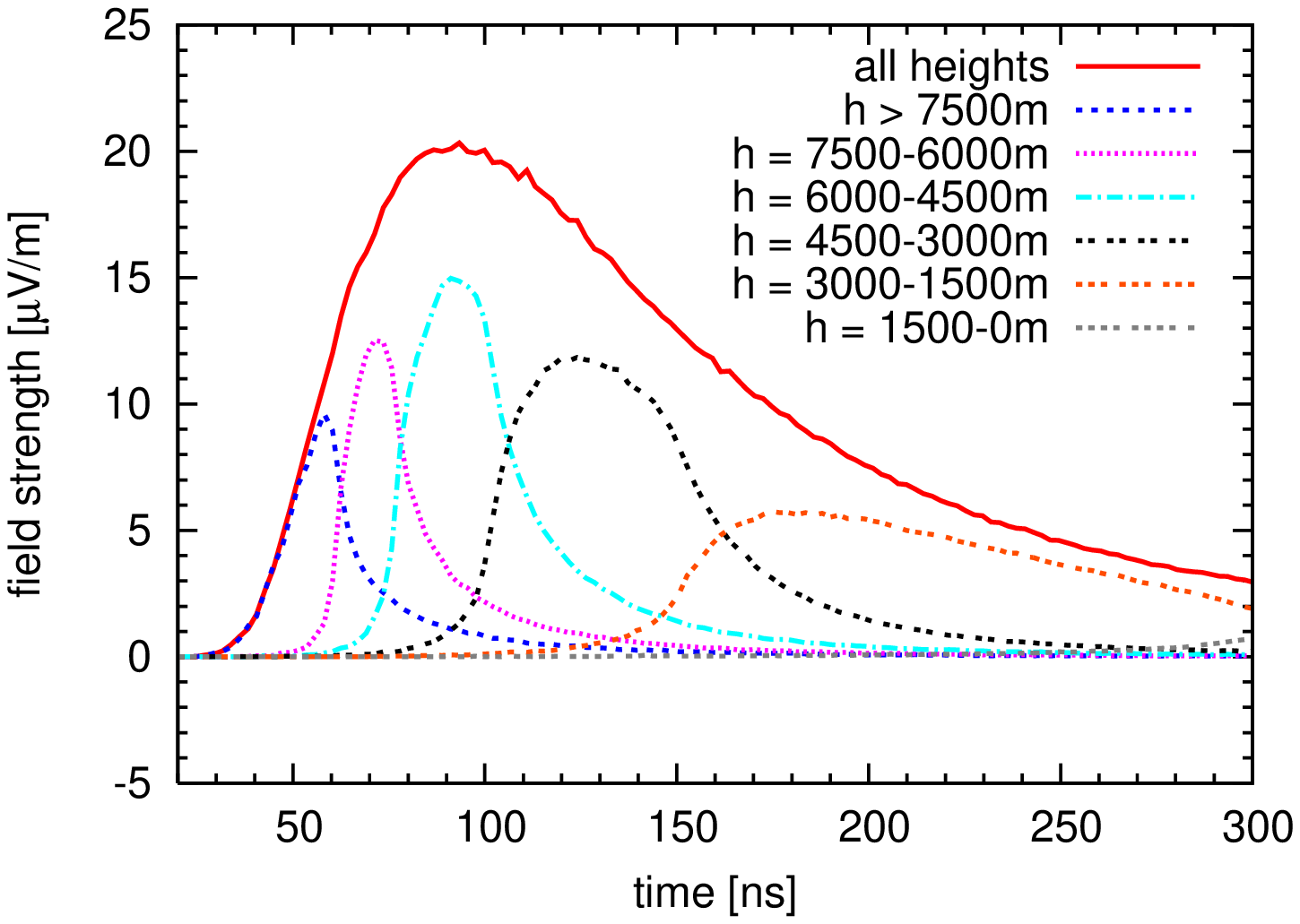}
\caption{\label{heightregimes525m}Contribution of different heights to the radio pulse at 525~m north from the shower centre.}
\end{minipage}
\end{figure}

Figures \ref{heightregimes75m} and \ref{heightregimes525m} illustrate how different regions of geometrical height above ground contribute to the radio pulses from the air shower. The qualitative behaviour is similar as for the stages of atmospheric depth, but with one important difference: the contributions from the different height regimes contribute approximately equally to the overall pulse. This, again, is explained by beaming effects and the fact that radio emission scales with the geometrical length of the particle tracks, not the atmospheric depth traversed.

\section{Conclusions}

We have analysed in detail how improving the air shower model from purely parametrised particle distributions to realistic histograms obtained with CORSIKA changes the radio emission predicted for a vertical $10^{17}$~eV proton-induced reference shower. The approach of keeping the well-understood radio code and only substituting the air shower model is ideally suited to understand the changes in detail.

The effects of the transition to realistic, histogrammed particle distributions have been evaluated step by step. One important change is introduced by the transition to the histogrammed arrival time distribution, which produces significantly narrower pulses and, consequently, flatter frequency spectra, in the shower centre region. (Measurements of radio pulse shapes close to the shower core in comparison with the simulations presented here could therefore be used to study the particle arrival time distributions in air showers.) As expected, by far the most important difference is caused by taking into account the realistic, much broader distribution of particle momentum angles to the shower axis. The histogrammed distribution dampens the radio emission significantly, as it causes a considerable fraction of the radiation to be emitted at large angles to the shower axis. It also became clear that the strong correlations between the different (e.g., energy, lateral and arrival time) distributions play an important role in shaping the radio emission and cannot be neglected. Another major improvement is the fact that the adoption of longitudinal shower evolution profiles obtained from CORSIKA allows us to analyse shower to shower fluctuations and differences related to the mass of the primary particle.

As a consequence of the much more realistic distributions in general, a qualitative change in the pulse shapes measured by observers along the east-west axis from the shower centre arises. Bipolarities in the pulses that were present in the simulations with parametrised air showers vanish when the realistic distributions are adopted. The bipolarities present in our earlier simulations seem to be artifacts of the over-simplified air shower model.

The track length parameter $\lambda$ was eliminated in a natural way by simulating long tracks with an appropriate number of independent, representative short tracks. This approach ensures that the particle distributions throughout the tracks always represent the local distributions and that ionisation losses along a track can be safely neglected. The radio pulses are slightly enhanced by this choice of $\lambda$, compensating for some of the damping introduced by the histogrammed air shower model.

Different modes of creating particles independently or in pairs, with equal numbers of electrons and positrons or the correct ratio, have been investigated. Taking into account the correct electron to positron ratio only slightly changes the emission along the east-west axis from the shower centre.

Another change is visible in the azimuthal distribution of the radio emission. The radio ``footprint'' becomes significantly more asymmetric, as the damping of the radio signal is more pronounced in the east and west than in the north and south. The radial decrease of the radio emission, however, stays approximately exponential, and the scale parameter does not change significantly either for the vertical shower studied here.

The new code allows us to analyse how different regimes identifiable in the shower particle distributions contribute to the radio pulses measured on the ground. In particular, different stages of the air shower evolution contribute to different time-windows of the radio pulses at moderate to large distances from the shower centre, where geometric time-delays play an important role. This shows that, in principle, information on the air shower evolution (such as the position of the shower maximum, which in turn is related to the mass of the primary particle) is encoded in the radio signal and could be accessible through radio measurements of cosmic ray air showers. The contributions of different slices in geometrical height above ground illustrate that beaming effects play an important role and that the geosynchrotron radio emission scales with the geometrical length of particle tracks, not the atmospheric depth traversed. This means that the same number of particles in a less dense medium produces more geosynchrotron radiation than in a dense medium. The analyses of the energy and radial distance regimes demonstrate that in the shower centre, high-energy particles located close to the shower axis dominate the emission, whereas at distances of a few hundred metres, lower-energy particles located further away from the shower axis dominate.

With REAS2, a next-generation, well-tested Monte Carlo code for the calculation of geosynchrotron radio emission from cosmic ray air showers is available. As a next step, we will repeat our earlier analysis of the influence of specific air shower parameters on the associated radio emission (cf.\ \citep{HuegeFalcke2005b}) with radio simulations based on CORSIKA-simulated showers. Additionally, we will assess the role of shower-to-shower fluctuations and the primary particle composition. In the future, we will incorporate the effect of the atmospheric refractive index profile in combination with the presence of systematic lateral shifts between the electron and positron distributions to take into account \v{C}erenkov-like radiation contributions.

\begin{ack}
The authors would like to thank T.\ Pierog and D.\ Heck for their outstanding support regarding many CORSIKA-related aspects of this work and the anonymous referee for several valuable comments that helped to improve the article. T.H.\ acknowledges very helpful discussions with S.\ Buitink, A.\ Haungs and H.\ Falcke. R.E.\ would like to thank N.\ Kalmykov and A.\ Konstantinov for many useful discussions.
\end{ack}

% \appendix

% \section{}
% \label{}

%________________________________________________________________________

% Bibliographic references with the natbib package:
% Parenthetical: \citep{Bai92} produces (Bailyn 1992).
% Textual: \citet{Bai95} produces Bailyn et al. (1995).
% An affix and part of a reference:
%   \citep[e.g.][Ch. 2]{Bar76}
%   produces (e.g. Barnes et al. 1976, Ch. 2).

%________________________________________________________________________

%\bibliography{references}
%\bibliographystyle{elsart-num}

\end{document}